\documentclass{aa}
\usepackage{graphicx}
\begin{document}
\def\los{\mathrel{\mathchoice {\vcenter{\offinterlineskip\halign{\hfil
$\displaystyle##$\hfil\cr<\cr\sim\cr}}}
{\vcenter{\offinterlineskip\halign{\hfil$\textstyle##$\hfil\cr
<\cr\sim\cr}}}
{\vcenter{\offinterlineskip\halign{\hfil$\scriptstyle##$\hfil\cr
<\cr\sim\cr}}}
{\vcenter{\offinterlineskip\halign{\hfil$\scriptscriptstyle##$\hfil\cr
<\cr\sim\cr}}}}}
\def\gos{\mathrel{\mathchoice {\vcenter{\offinterlineskip\halign{\hfil
$\displaystyle##$\hfil\cr>\cr\sim\cr}}}
{\vcenter{\offinterlineskip\halign{\hfil$\textstyle##$\hfil\cr
>\cr\sim\cr}}}
{\vcenter{\offinterlineskip\halign{\hfil$\scriptstyle##$\hfil\cr
>\cr\sim\cr}}}
{\vcenter{\offinterlineskip\halign{\hfil$\scriptscriptstyle##$\hfil\cr
>\cr\sim\cr}}}}}

   \title{Chemistry as a probe of the 
          structures and evolution of massive star-forming regions
	  } 

   \author{S. D. Doty \inst{1}
           \and
	   E. F. van Dishoeck \inst{2}
	   \and
           F. F. S. van der Tak \inst{2,3}
	   \and
	   A. M. S. Boonman \inst{2}
          }

   \offprints{S. Doty}

   \institute{Department of Physics and Astronomy,
              Denison University, Granville, OH  43023
          \and
              Sterrewacht Leiden, P.O. Box 9513,
              2300 RA Leiden, The Netherlands
          \and
	      Max-Planck-Institut f\"ur Radioastronomie,
	      Auf dem H\"ugel 69, 53121 Bonn, Germany 
             }

   \date{Received; accepted}

   \titlerunning{Chemistry in AFGL 2591}

   \abstract{
   We present detailed thermal and gas-phase chemical models
   for the envelope of the massive star-forming region
   AFGL 2591.  By considering both time- and space-dependent
   chemistry, these models are used to study both the  
   physical structure proposed by van der Tak
   et al. (\cite{vdt1999}; \cite{vdt2000}), as well as 
   the chemical evolution of this region.  
   The model predictions are compared with observed 
   abundances and column densities for 29 species.
   The observational data cover a wide range of
   of physical conditions within the source, but significantly
   probe the inner regions where interesting high-temperature
   chemistry may be occurring.
   Taking appropriate care when comparing
   models with both emission and absorption measurements, 
   we find that the majority of the chemical structure can
   be well-explained.  In particular, we find that the 
   nitrogen and hydrocarbon chemistry can be significantly
   affected by temperature, with the possibility of 
   high-temperature pathways to HCN.  While we cannot
   determine the sulphur reservoir, the observations can be
   explained by models with the majority of the sulphur in 
   CS in the cold gas, SO$_{2}$ in the warm gas, and 
   atomic sulphur in the warmest gas.  Because the model
   overpredicts CO$_{2}$ by a factor of 40, 
   various high-temperature destruction mechanisms are explored, including
   impulsive heating events.
  The observed abundances of ions such as HCO$^+$ and N$_2$H$^+$
   and the cold gas-phase production of HCN  
   constrain the cosmic-ray
   ionization rate to $\sim 5.6 \times 10^{-17}$ s$^{-1}$,
   to within a factor of three.
   Finally, we find that the model and observations can
   simultaneously agree at a reasonable level and
   often to within a factor of three
   for $7 \times 10^{3} \leq t(\mathrm{yrs}) \leq 5 \times 10^{4}$,
  with a strong preference for $t \sim 3 \times 10^{4}$ yrs since
the collapse and formation of the central luminosity source.
   }
   
   \maketitle

%

\section{Introduction}
The distribution and composition of dust and gas around isolated 
low-mass young stellar objects (YSOs) has been well-studied
both observationally and theoretically.  Unfortunately,
much less is known about the distribution and composition
of material around high-mass YSOs (see e.g., Churchwell 
\cite{churchwell93}, \cite{churchwell99}).
The higher densities and masses, and shorter
lifetimes associated with massive star formation 
suggest that differences between regions of high- and low-mass
star formation can be expected.  

Recent observational advances (e.g., submillimeter beams
of $\sim 15''$ sampling smaller regions of higher critical densities,
interferometry at 1 and 3 mm, and ground- and
space-based infrared observations of gas and ices)
have led to a new and better understanding of the
environment around massive YSOs (see e.g., Garay \& Lizano \cite{gl99},
van Dishoeck \& Hogerheijde 1999, Hatchell et al. \cite{hatchell2000},
Beuther et al. \cite{beuther2002}).
In this vein, van der Tak et al. (\cite{vdt1999}, \cite{vdt2000}) 
have conducted detailed
multi-wavelength studies of high-mass YSOs, and begun to form a picture
for the physical structure of some of these regions.
The proposed material distributions in the envelopes
fit a wide variety of continuum and spectral
line data.  However, they are incomplete without a detailed
thermal and chemical structure.  The proposed material distribution
can be used to test the chemical structure and evolution of
the envelope, and the combined results can eventually be used to 
compute line strengths and profiles for direct
comparison with observations.

Significant work has been 
involved in developing an understanding of the chemistry
of star-forming regions.  This ranges from
studies of cold, dark clouds 
(e.g., Herbst \& Klemperer \cite{herbstklemperer},
Prasad \& Huntress \cite{prasadhuntress},
Leung, Herbst, \& Huebner \cite{leungherbsthuebner},
Gwenlan et al. \cite{gwenlanetal}) 
to ``hot cores''
(see e.g., reviews by Millar \cite{millarreview1993}; 
Walmsley \& Schilke \cite{walmsleyschilke1992};
Kurtz et al. \cite{kurtzetal2000}).  In nearly all cases, however, 
the chemistry is considered for a homogeneous cloud,
or a point within a cloud (see, though, Xie et al. \cite{xieetal1995}
and Bergin et al. \cite{berginetal1995} for counter examples).
Unfortunately, the physical conditions (i.e., temperature and density)
vary strongly with position within the envelope,
meaning that potentially extreme chemical variations may occur 
between the source center and the observer.
It is this strong variation of chemical composition with position and time
that may provide one of the best benchmarks of our
understanding of both the structures and evolution of massive 
star-forming regions.

In this paper, we utilize position-dependent thermal balance 
and time- and position-dependent chemical modeling to probe the validity of
the physical structures proposed by van der Tak et al.  
(\cite{vdt1999} \& \cite{vdt2000}), and more importantly, to
study the chemical evolution of AFGL 2591.
In particular, taking their structure as a starting point,
we construct detailed models for the gas-phase chemistry
of this source, and compare the results with observations.  
AFGL 2591 is a massive ($\sim 42 M_{\odot}$
within $r=3\times10^{4}$ AU), luminous ($\sim 2 \times 10^{4} L_{\odot}$)
infrared source with many of the properties thought to characterize YSOs.
While most massive stars form in clusters, AFGL 2591 has the
advantage that it is forming in relative isolation -- 
allowing us to study its physical, thermal, and chemical
structures without influence from other nearby massive sources.
It has the further advantage of being well-observed both in the
continuum and in a variety of molecular lines.   

This paper is organized as follows.  The existing observations
providing the model constraints are briefly discussed  
in Section 2.  In Section 3, the model is described.
The model is then applied to AFGL 2591 and compared with
the observational results in Section 4.  In Section 5, we 
compare our time-dependent model predictions with the observations
in order to constrain the chemical age of the envelope.
Finally, we summarize our results and conclude in Section 6.

\section{Existing observations and usage}
AFGL 2591 has been well-observed both in the continuum and in
various molecular lines.  While no new observations are 
presented in this paper, it is important to briefly note and
discuss the relevant observations as they provide the
constraints placed on the model.  

\subsection{Continuum}
AFGL 2591 has been observed in the range $2-60000$ ${\mu}$m by
Lada et al (\cite{ladaetal}), Aitken (\cite{aitkenetal}), 
Sandell (1998, private communication), and van der
Tak et al. (\cite{vdt1999}). 
These results were analyzed by van der Tak et al. (\cite{vdt2000} -- see 
Sect. 3 below) 
to constrain the density distribution and grain properties --
necessary for not only the thermal structure, but also to
properly evaluate
the gas thermal balance and hence obtain the gas temperature as
a function of position.

\subsection{Molecular lines}
A wide variety of observations, both in the infrared and
submillimeter, have been conducted of molecular gas in
AFGL 2591, some of which are as of yet unpublished.  
The results are summarized in  
Table \ref{molecularobs}, where the species, observed 
abundance [$x(\mathrm{X}) \equiv n(\mathrm{X})/n(\mathrm{H}_{2})$] or
column density [$N(\mathrm{X})$], inferred excitation temperature,
method of analysis, weight used in selecting the most
important of the relevant observations, type of observation, and 
reference are listed.

\subsection{Notes on Table \ref{molecularobs} and usage of the data}
The observation type is listed in Table \ref{molecularobs}
as this is significant for
comparing the results with observations.  For
infrared absorption lines, the molecules observed are along
the (narrow) line of sight to the 
background continuum source.  Consequently, these 
results should be compared to model ``radial column densities'', 
namely 
$N_{\mathrm{radial}} \equiv \int n(r) \mathrm{d}r$.  
On the other hand, submillimeter emission lines 
arise from throughout the envelope.  In these cases, averages over
the beam are used in comparing predicted and observed
column densities.  Here, the ``beam-averaged column density''
is defined as 
$N_{\mathrm{beam}} \equiv
\int \int n(z,p) ~ \mathrm{d}z ~ G(p) 2 \pi p ~ \mathrm{d}p /
\int G(p) 2 \pi p ~ \mathrm{d}p$ where $p$ is the 
impact parameter, and $G(p)$ is the beam response function. 
We also divide the data in this 
fashion, as we expect many of the uncertainties in the
analysis to be similar for one type of observation.

   \begin{table*}
      \caption[]{Inferred Column Densities and Abundances Toward AFGL 2591}
         \label{molecularobs}
         \begin{tabular}{lrrrcccl}
            \hline
      Molecule  & $x$ & $N$ (cm$^{-2}$) & $T_{\mathrm{ex}}$(K)  & Method & Weight & Data & Ref\\
            \hline
      H$_{2}$        & {} & $9.6(22)$ & {-} & Scale $N(\mathrm{CO})$   & 2  & {-} & a  \\
      HCN$^{(\alpha 1)}$  & $\sim 1(-8)$ & {} & $\los 230$ & NLTE RT Model & 3  & submm - JCMT & b \\
      {}$\ldots^{(\alpha 2)}$   & $\sim 1(-6)$ & {} & $\gos 230$ & NLTE RT Model & 3  & submm - JCMT & b \\
      {}HCN$^{(\alpha 1)}$   & {} & $4.0(16)$ & 600 & Absn. Depth & 2 & IR - ISO & a \\
      {}HCN$^{(\beta 1)}$    & {} & $2.0(15)$ & 38 (CO) & Absn. Depth & 2 & IR - ISO & a \\
      {}$\ldots^{(\beta 2)}$ & {} & $4.5(16)$ &1010 (CO)& Absn. Depth & 2 & IR - ISO & a \\
      {}HCN$^{(\alpha 1)}$   & {} & $\leq 1.7(16)$ &38 (CO)& Absn. Depth & 2  & IR - IRTF & c \\
      {}$\ldots^{(\alpha 2)}$& {} & $2.0(16)$ &200 (CO) & Absn. Depth & 2 & IR - IRTF & c \\
      {}$\ldots^{(\alpha 3)}$& {} & $1.6(16)$ &1010 (CO) & Absn. Depth & 2 & IR - IRTF & c \\
      HNC$^{(\alpha 1)}$     & {} & $2.9(13)$ & {-} & NLTE / escape prob. & 2 & submm - JCMT & d \\
      HNC$^{(\alpha 1)}$ & $1.0(-8)$ & {} & {-} & NLTE RT Model & 3 & submm - JCMT & n \\
      HC$_{3}$N$^{(\alpha 1)}$  & {} & $5.0(12)$ & {-} & NLTE / escape prob. & 2 & submm - JCMT or CSO & d \\
      HC$_{3}$N$^{(\alpha 1)}$ & $2.0(-8)$ & {} & {-} & NLTE RT Model & 2 & submm - JCMT & n \\
      HCO$^{+}${}$^{(\alpha 1)}$ & $1.0(-8)$ & {} & {-} & NLTE RT Model & 3 & submm - JCMT & e \\
      HCS$^{+}${}$^{(\alpha 1)}$ & $3.0(-10)$ & {} & {-} & NLTE RT Model & 3 & submm - JCMT & n \\
      H$_{3}^{+}${}$^{(\alpha 1)}$ & {} & $1-3(14)$ & {-} & Absn. Depth & 2 & IR - UKIRT/IRTF & f \\
      H$_{2}$O$^{(\alpha 1)}$  & {} & $3.5(18)$ & $450$ & Absn. Depth & 2 & IR - ISO   & g  \\
      H$_{2}$S$^{(\alpha 1)}$  & {} & $\leq 1.0(19)$ & {-} & Absn. Depth & 2 & IR - ISO  & h \\
      H$_{2}$CO$^{(\alpha 1)}$ & $2.0(-9)$ & {} & {-} & NLTE RT Model & 3 & submm - JCMT & i \\
      {}H$_{2}$CO$^{(\beta 1)}$  &{} & $8.0(13)$ & 89 & LTE Rot. Diagram & 2 & submm - JCMT & e \\ 
      {}H$_{2}$CS$^{(\alpha 1)}$ & $1.0(-9)$ & {} & {-} & NLTE RT Model & 2 & submm - JCMT & n \\
      CI$^{(\alpha 1)}$     & {} & $\leq 6.8(17) $ & {-} & NLTE / escape prob. & 2 & submm - CSO & j \\
      C$^{+}${}$^{(\alpha 1)}$  & {} & $\leq 6.8(17)$ & {-} & LTE escape prob. & 2 & IR - ISO & k \\ 
      C$_{2}$H$^{(\alpha 1)}$ & $2.0(-9)$ & {} & {-} & NLTE RT Model & 3 & submm - JCMT & n \\
      C$_{2}$H$_{2}^{(\alpha 1)}$ & {} & $\leq 2.0(16)$ & 900 & Absn. Depth & 2 & IR - ISO & a \\
      {}C$_{2}$H$_{2}^{(\beta 1)}$  & {} & $\leq 1.0(15)$ & 38 (CO) & Absn. Depth & 2 & IR - ISO & a \\
      {}$\ldots^{(\beta 2)}$  & {} & $2.0(16)$ & 1010 (CO) & Absn. Depth & 2 & IR - ISO & a \\
      {}C$_{2}$H$_{2}^{(\alpha 1)}$ & {} & $\leq 8.0(14)$ & 38 (CO) & Absn. Depth & 2 &  IR - IRTF & c \\
      {}$\ldots^{(\alpha 2)}$ & {} & $4.2(15)$     &200 (CO) & Absn. Depth & 2 & IR - IRTF & c \\
      {}$\ldots^{(\alpha 3)}$ & {} & $1.0(16)$     &1010 (CO) & Absn. Depth & 2 & IR - IRTF & c \\
      CH$_{4}^{(\alpha 1)}$ & {} & $2.5(17)$ & $ \geq 1000$ & Absn. Depth & 2 & IR - ISO & h\\                          
      {}CH$_{4}^{(\alpha 1)}$ & {} & $\leq 8.0(15)$ & 38 (CO) & Absn. Depth & 2 & IR - IRTF & c \\
      {}$\ldots^{(\alpha 2)}$ & {} & $\leq 1.0(17)$ &200 (CO) & Absn. Depth & 2 & IR - IRTF & c \\
      {}$\ldots^{(\alpha 3)}$ & {} & $\leq 1.3(18)$ &1010 (CO) & Absn. Depth & 2 & IR - IRTF & c \\
      CH$_{3}$OH$^{(\alpha 1)}$  & $2.6(-9)$ & {} & $\leq 90$ & NLTE RT Model & 3 & submm - JCMT & i \\
      $\ldots^{(\alpha 2)}$  & $8.0(-8)$ & {} & $\geq 90$ & NLTE RT Model & 3 & submm - JCMT & i \\
      CH$_{3}$OH$^{(\beta 1)}$   & {} & $1.2(15)$ & 163 & rot. diagram & 2 & submm - JCMT & i \\
      CH$_{3}$CN$^{(\alpha 1)}$ & $2.0(-8)$ & {} & {-} & NLTE RT Model & 2 & submm - JCMT & n \\
      \hline
      \end{tabular} \\
      \end{table*}

   \begin{table*}
      \addtocounter{table}{-1}
      \caption[]{Inferred Column Densities and Abundances Toward AFGL 2591
                 (continued from last page)}
         \label{molecularobs2}
         \begin{tabular}{lrrrcccl}
            \hline
      Molecule  & $x$ & $N$ (cm$^{-2}$) & $T_{\mathrm{ex}}$(K)  & Method & Weight & Data & Ref\\
            \hline
      CO$^{(\alpha 1)}$ & {} & $1.3(19)$ & {-} & Absn. Depth & 2 & IR - CFHT & o  \\
      {}CO$^{(\alpha 1)}$ & {} & $3.4(19)$ & {-} & NLTE RT Model & 3 & submm - JCMT & e \\
      CO$_{2}^{(\alpha 1)}$ & {} & $2.5(16)$ & 500 & Absn. Depth & 2 & IR - ISO & g    \\
      CS$^{(\alpha 1)}$ & $3.0(-9)$ & {} & 40 & NLTE RT Model & 3 & submm - JCMT & e \\
      {}CS$^{(\alpha 1)}$ & {} & $\leq 2.6(15)$ &38 (CO) & Absn. Depth & 2 & IR - IRTF & c\\
      {}$\ldots^{(\alpha 2)}$ & {} & $\leq 3.4(15)$ &200 (CO) & Absn. Depth & 2 & IR - IRTF & c\\
      {}$\ldots^{(\alpha 3)}$ & {} & $\leq 9.0(15)$ &1010 (CO) & Absn. Depth & 2 & IR - IRTF &c \\
      {}CN$^{(\alpha 1)}$ & $5.0(-8)$ & {} & {-} & NLTE RT Model & 2 & submm - JCMT & n \\
      OH$^{(\alpha 1)}$ & {} & $\geq 4.7(14)$ & {-} & Absn. Depth & 2 & IR - ISO & h \\
      O$_{2}^{(\alpha 1)}$ & $\leq 1.0(-6)$ & {} & {-} & NLTE / opt. thin & 3 & submm - SWAS & l \\
      OCS$^{(\alpha 1)}$   & {} & $1.0(14)$ & {-} & NLTE / escape prob. & 2 & submm - JCMT & d \\
      OCS$^{(\alpha 1)}$ & $4.0(-8)$ & {} & {-} & NLTE RT Model & 3 & submm - JCMT & n \\ 
      NH$_{3}^{(\alpha 1)}$& {} & $\leq 5.0(14)$ &38 (CO) & Absn. Depth & 2 & IR - IRTF & c \\
      {}$\ldots^{(\alpha 2)}$  & {} & $\leq 1.0(15)$ & 200 (CO) & Absn. Depth & 2 & IR - IRTF & c \\
      {}$\ldots^{(\alpha 3)}$  & {} & $\leq 7.0(15)$ &1010 (CO) & Absn. Depth & 2 & IR - IRTF & c \\
      NH$_{3}^{(\alpha 1)}$ & $2.0(-8)$ & {} & {-} & NLTE RT Model & 2 & cm - Effelsberg & n \\
      N$_{2}$H$^{+}${}$^{(\alpha 1)}$ & {} & $1.4(12)$ & {-} & NLTE / escape prob. & 2 & submm - JCMT & d \\
      N$_{2}$H$^{+}${}$^{(\alpha 1)}$ & $5.0(-10)$ & {} & {-} & NLTE RT Model & 3 & submm - JCMT & n \\
      SO$^{(\alpha 1)}$ & $2.0(-8)$ & {} & {-} & NLTE RT Model & 3 & submm - JCMT & n \\
      SO$_{2}^{(\alpha 1)}$ & {} & $6.0(16)$ & 200 & Absn. Depth & 2 & IR - ISO & m \\

            \hline
         \end{tabular} \\
$^{\mathrm{ }}$ $a(b)$ means $a \times 10^{b}$ \\
In {}$^{(\alpha 1)}$ the first symbol
denotes the fit number ($\alpha$ is the first fit,
$\beta$ is the second, $\ldots$), and the second is the component
of that fit (1 is the first component, 2 is the second, $\ldots$). \\
The (CO) notation signifies that $T_{\mathrm{ex}}$ was forced to be one
of the three CO temperatures from Mitchell et al. \cite{mitchell89}\\
Method \& Weight:  The method used to infer, and the significance we 
ascribe to, the observational result (higher is better).\\
$^{\mathrm{a}}$ Lahuis \& van Dishoeck \cite{Lv}, 
$^{\mathrm{b}}$ Boonman et al. \cite{boonmanetal2001}, 
$^{\mathrm{c}}$ Carr et al. \cite{CELZ}, 
$^{\mathrm{d}}$ van Dishoeck 2001 (private communication), 
$^{\mathrm{e}}$ van der Tak et al. \cite{vdt1999}, 
$^{\mathrm{f}}$ McCall et al. \cite{mccalletal},
$^{\mathrm{g}}$ Boonman et al. \cite{boonmanetal2000}, 
$^{\mathrm{h}}$ Boonman 2001 (private communication), 
$^{\mathrm{i}}$ van der Tak et al. \cite{vdt2000}, 
$^{\mathrm{j}}$ Choi et al. \cite{choietal}, 
$^{\mathrm{k}}$ Wright 2001 (private communication), 
$^{\mathrm{l}}$ Goldsmith et al. \cite{swaso2}, 
$^{\mathrm{m}}$ Keane et al. \cite{keaneetal},
$^{\mathrm{n}}$ van der Tak 2002 (in preparation),
$^{\mathrm{o}}$ Mitchell et al. \cite{mitchell89}
   \end{table*}

In columns 2 and 3 of Table \ref{molecularobs} we list the 
inferred fractional abundance or column density
of the given molecule toward AFGL 2591.  This is done to 
provide the most comprehensive set of information with
which to compare our models.  

While determination of column density is relatively straightforward
for infrared absorption lines in the limit of no re-emission,
the situation is more difficult for emission lines as the
emission may arise from a range of radii, and thus a range of
densities, temperatures, chemical abundances, and optical depths.
To combat this, some effort has been made recently to 
determine the fractional abundance within the envelope 
through detailed, non local thermodynamic equilibrium (NLTE)
radiative transfer (RT) modeling (van der Tak et al. \cite{vdt1999}).  
When this is
done, we view the inferred abundances as superior to
pure column densities as they account for many of the
potential errors in determining the column density.  As examples,
van der Tak, van Dishoeck, \& Caselli (\cite{vvc}) and Boonman et al.
(\cite{boonmanetal2001}) 
have used such modeling to suggest ``jump'' models for
the chemical enhancement of species within certain regions of YSO envelopes.  
As a result, in column 5 we note the method used in determining the
observational result.  We also use these criteria to assign
a weight (higher is better) in column 6 to denote which
data/fits we view as superior.  In cases where the fit due
to radiative transfer modeling is only moderate, we give this
result the same weight as the results from other methods.

In both cases, where an excitation temperature can be assigned to the
data, we note the temperature for that component as $T_{\mathrm{ex}}$
in column 4 of Table \ref{molecularobs}.  While $T_{\mathrm{ex}}$ is
not necessarily equal to the kinetic temperature, it does give some
indication as to the region from which the observation arises. The
values of 38, 200 and 1010~K refer to the excitation temperatures of
CO found by Mitchell et al.\ (1989) in infrared absorption line
studies. The 200~K component is thought to be associated with shocked
outflowing material, whereas the other two temperatures refer to the
quiescent envelope.

Finally, we note the relative importance of different
measurements for probing various regions in the envelope.
Absorption is confined to the narrow line of 
sight toward the central source.  For centrally-condensed envelopes,
the column density is dominated by the interior.  This makes
absorption measurements useful for probing the warm interior.
On the other hand, emission measurements can and often do arise
from throughout the envelope.  When the density falls off slower
than $r^{-2}$ as is the case for AFGL 2591 (van der Tak et al.
\cite{vdt1999}) 
the outer portion of the envelope dominates the mass, and
so emission measurements are often more useful for probing the 
cool exterior.   These expectations are relatively consistent with 
the results of Table \ref{molecularobs}, where many of the 
absorption measurements include significant high excitation temperature
components, while the inferred excitation temperatures for the
emission data are generally much lower.

\section{Model}

In this section, a brief synopsis of the physical, thermal, 
and chemical models are provided.  For more detailed information, see
van der Tak et al. (\cite{vdt1999}, \cite{vdt2000}), 
Doty \& Neufeld (\cite{DN}), and references therein.
For reference, the model parameters are reproduced in 
Table \ref{modelparameters}. 

\subsection{Physical model}
Our model for AFGL 2591 concentrates on the
extended envelope of source.  While an inner disk may be 
present, OVRO interferometric observations by
van der Tak et al. (\cite{vdt1999}) suggest an unresolved
central source of radius $30<r\mathrm{(AU)}<1000$.
These and other continuum observations
were analyzed by van der Tak et al. (\cite{vdt2000}) using
a modified version of the self-consistent continuum radiative
transfer model of Egan, Leung, \& Spagna (\cite{eganetal}).
Based upon the fit to the continuum flux and surface brightness,
as well as CS line data, they constrained the density to a power law
of the form $n(r)=n_{0}(r_{0}/r)^{\alpha}$.  In particular,
they found a best fit with ${\alpha}=1.0$, and 
$n_{0} \equiv n(\mathrm{H}_{2},r=r_{0}=2.7\times 10^{4} \mathrm{AU}) 
= 5.8 \times 10^{4}$ cm$^{-3}$.  We adopt these values for the 
remainder of the paper.  Our inner radial position was chosen
to be $r_{\mathrm{in}} = 2 \times 10^{2}$ AU, corresponding to $T \sim 440$ K
(see below).  This inner radius was chosen not only for
consistency with the observations of van der Tak et al. (\cite{vdt2000}),
but also as extrapolation of a density power law
further into the interior of the envelope
leads to column densities inconsistent with
observations (see Sect. 4.2 below).

Following the analyses of van der Tak et al. (\cite{vdt1999},
\cite{vdt2000}) and Doty \& Neufeld (\cite{DN}), we assume that the physical
and thermal model does not change significantly with time
so that an equilibrium may be achieved, but we do allow
for a time-dependent chemical evolution.  
While the collapse and rise in
luminosity will occur on short time scales ($< 1000$ yr), it is the combination
of density structure and luminosity of the central source
that sets the temperature
structure.  Therefore, as long as the envelope mass and
luminosity do not significantly change, we can consider the source as
approximately constant over the $\sim 10^{5}$ yrs in which the
envelope will be dissipated (Hollenbach et al. \cite{holl94},
Richling \& Yorke \cite{richlingyorke}).  
To see this, consider the
fact that the free fall and sound-crossing times at the outer
edge are both $\sim 2 \times 10^{5}$ yrs.  While these timescales
are smaller closer to the center, accretion events should probably only
be important in the very interior.

Finally, we note that an outflow has been observed toward this
source (see, e.g., Bally \& Lada \cite{ballylada}, 
Mitchell et al. \cite{mitchell89}).  
However, spectroscopy shows that nearly all 
submillimeter lines with
the exception of CO can be assigned to the envelope as their linewidths
are only $\sim$ few km s$^{-1}$ 
(see, e.g., van der Tak et al. \cite{vdt1999},
and recent and upcoming infrared data from TEXES by 
Knez et al. \cite{knezaas} and Boonman et al. -- in preparation).  
Only CO has a significant fraction
of the observed material in the outflow.  This assignment of material
to the envelope rather than the outflow is also justified 
{\em a posteriori} as our models are able to reproduce a good
deal of the observed chemistry without the requirement
of shock chemistry. Because the submillimeter lines probe
high excitation gas, the lower density surrounding cloud is
automatically filtered out.

%
   \begin{figure}
      \resizebox{\hsize}{!}{\includegraphics{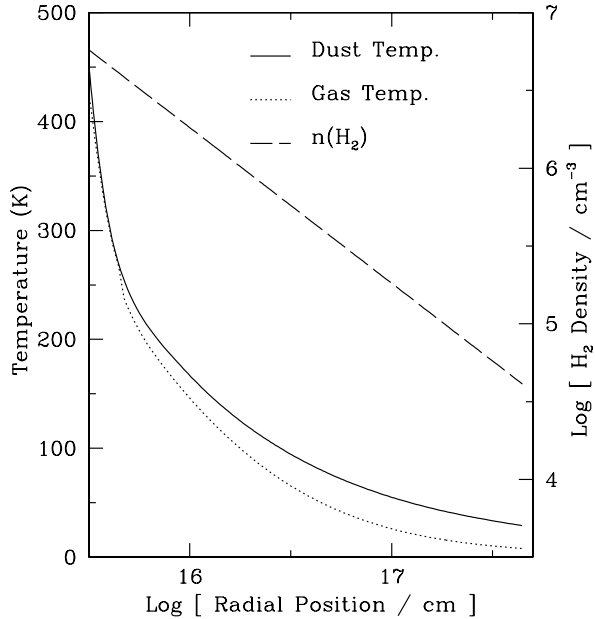}}
      \caption[]{Physical and thermal structure of AFGL 2591.
                 The density and dust results from the model of van der
		 Tak et al. (\cite{vdt2000}). 
                 The gas temperatures are calculated from the 
		 detailed thermal balance,
                 similar to Doty \& Neufeld (\cite{DN}).  Note that 
                 $T_{\mathrm{gas}} \sim T_{\mathrm{dust}}$.
              }
         \label{tempdens}
   \end{figure}
%

\subsection{Thermal model}
The equilibrium gas temperature within the cloud is determined
by the balance between heating and cooling.  The
gas heating is dominated by gas-grain collisions, and the
dust temperature is determined from the self-consistent
solution to the continuum radiative transfer problem as above.
The Neufeld, Lepp, \& Melnick (\cite{NLM}) cooling
functions were adopted, with modifications as noted in Doty \& Neufeld
(\cite{DN}). 
Furthermore, as the Neufeld, Lepp, \& Melnick (\cite{NLM}) 
cooling functions were constructed
assuming a singular isothermal sphere (with a commensurate
$n \propto r^{-2}$ density power law), they were modified to
be applicable to the $r^{-1}$ power law adopted here. 
This entailed two corrections.  First, the column densities
had to be computed correctly at each position, rather than
simply relying upon the local density.  Second, the
cooling functions for the tabular results were modified by
a factor $[N({\alpha}=2)/N({\alpha})]^{f}$, where 
$N({\alpha}) \equiv \int_{r_{\mathrm{out}}}^{r}n(r)~\mathrm{d}r$ 
is the column density for a power law distribution
$r^{-{\alpha}}$.  Here $f$ varies linearly with $\log (N)$
from $-0.5$ at $N=10^{16}$ cm$^{-2}$ to 
$-1.0$ at $N=10^{21}$ cm$^{-2}$.  We take $f=0$ for $N<10^{16}$ cm$^{-2}$, 
and $f=-1.0$ for $N>10^{21}$ cm$^{-2}$.  This factor is chosen to match the
functional dependence of the cooling rate on the column 
density as described in Neufeld, Lepp, \& Melnick (\cite{NLM})
for H$_2$O -- the dominant 
coolant utilized in tabular form -- and is consistent with
the fact that the cooling rate should be inversely proportional
to the column density for opaque sources.  The resulting gas temperature
distribution is shown in Fig. \ref{tempdens}, and 
is physically similar to that of Doty \& Neufeld (\cite{DN}), namely 
that $T_{\mathrm{gas}} \sim T_{\mathrm{dust}}$, as was
assumed by van der Tak et al. (\cite{vdt1999}, \cite{vdt2000}).  
For comparison, models run assuming 
$T_{\mathrm{gas}}=T_{\mathrm{dust}}$ show no significant
differences.

   \begin{table}
      \caption[]{Model Parameters}
         \label{modelparameters}
         \begin{tabular}{lll}
            \hline
            Parameter & Value & Ref. \\ 
            \hline
     Outer radius (AU) & $3.0(4)$ & a \\
     Inner radius (AU) & $2.0(2)$ & a \\
     Density $[n(r)=n_{0}(r_{0}/r)^{\alpha}]$ & {} & {} \\
     $\ldots$ Exponent $[{\alpha}]$ & $1.0$ & a,b \\
     $\ldots$ Ref. position $[r_{0}]$ (AU) & $2.7(4)$ & b \\
     $\ldots$ Ref. H$_{2}$ density $[n_{0}]$ (cm$^{-3}$) & $5.8(4)$ & b \\
     CR ionization rate $[\zeta]$ (s$^{-1}$)& $5.6(-17)$ & c \\
     {} & {} & {} \\
     Initial Abundance ($T>100\mathrm{K}$) & {} & {} \\
     CO & $3.7(-4)$ & a \\
     CO$_{2}$ & $3.0(-5)$ & d \\
     H$_{2}$O & $1.5(-4)$ & d \\
     H$_{2}$S & $1.6(-6)$ & see text \\
     N$_{2}$  & $7.0(-5)$ & e \\
     CH$_{4}$ & $1.0(-7)$ & e \\
     C$_{2}$H$_{4}$ & $8.0(-8)$ & e \\
     C$_{2}$H$_{6}$ & $1.0(-8)$ & e \\
     OI & $0.0(0)$ & e \\ 
     H$_{2}$CO & $1.2(-7)$ & e \\
     CH$_{3}$OH & $1.0(-6)$ & e \\
     S  & $0.0(0)$ & e \\
     Fe & $2.0(-8)$ & e \\
     {} & {} & {} \\
     Initial Abundance ($T<100\mathrm{K}$) & {} & {} \\
     CO & $3.7(-4)$ & a \\
     CO$_{2}$ & $0.0(0)$ & f \\
     H$_{2}$O & $0.0(0)$ & f \\
     H$_{2}$S & $0.0(0)$ & f \\
     N$_{2}$  & $7.0(-5)$ & e \\
     CH$_{4}$ & $1.0(-7)$ & e \\
     C$_{2}$H$_{4}$ & $8.0(-8)$ & e \\
     C$_{2}$H$_{6}$ & $1.0(-8)$ & e \\
     OI & $8.0(-5)$  & g \\
     H$_{2}$CO & $0.0(0)$ & f \\
     CH$_{3}$OH & $0.0(0)$ & f \\
     S  & $6.0(-9)$  & see text \\ 
     Fe & $2.0(-8)$ & e \\
            \hline
     \end{tabular}\\
$^{\mathrm{ }}$ { } $a(b)$ means $a \times 10^{b}$, 
$^{\mathrm{ }}$ All abundances are gas-phase, and relative to H$_{2}$ \\
$^{\mathrm{a}}$ van der Tak et al. \cite{vdt1999}, 
$^{\mathrm{b}}$ van der Tak et al. \cite{vdt2000}, 
$^{\mathrm{c}}$ van der Tak \& van Dishoeck \cite{vv}, 
$^{\mathrm{d}}$ Boonman et al. \cite{boonmanetal2000}, 
$^{\mathrm{e}}$ Charnley \cite{C97}, 
$^{\mathrm{f}}$ assumed frozen-out or absent in cold gas-phase, 
$^{\mathrm{g}}$ taken to be $\sim$ consistent with Meyer, Jura, \& Cardelli 1998 
   \end{table}

\subsection{Chemical model}
The chemical model is based upon the UMIST 
gas-phase chemical reaction network 
(Millar, Farquhar, \& Willacy \cite{MFW}). 
Using this network, we construct pseudo time-dependent models
of the evolution of the chemical abundances.  We do this 
over a range of 30 radial grid points, providing a 
time- and space-dependent chemical evolution.  The 
local parameters (hydrogen density, temperature,
and optical depth) at each radial point are taken from the 
physical and thermal structure calculations above.
For our initial abundances, we follow Charnley 
(\cite{C97}; private communication).
These parameters allow us to reproduce  
many of the results of the hot core models of Charnley
(\cite{C97}; private communication), with most discrepancies 
directly attributable to differences in adopted reaction rates.   

We also include the approximate effects of freeze-out onto 
dust grains by initially depleting certain species below 100 K (see
Sect. 4.7 for discussion of H$_{2}$CO and CH$_{3}$OH).  We 
attempt to minimize this effect by predominantly depleting those
species that have high observed solid-phase abundances.
Our initial fractional abundances relative to H$_{2}$, 
as well as other model parameters are listed in Table \ref{modelparameters}.

The cosmic-ray ionization rate is taken from 
van der Tak \& van Dishoeck (\cite{vv}) for AFGL 2591,
and will be discussed in Sect. 4.5. 
The effects of cosmic-ray induced photochemistry were ignored.
The initial sulphur abundance was chosen to make the models
agree with observations (see Sect. 4.6).  
The assumed sulphur abundances are in general agreement with observations
for both the warm (e.g., toward Orion by Minh et al. \cite{minhetal1990}),
and the cold (e.g., Irvine, Ohishi, \& Kaifu \cite{irvineetal1991}) 
components.

The effects of photodissociation from the ISRF at the outer 
boundary are included, but are generally small due to the 
high optical depth, and the coarseness of the
spatial grid considered.

\section{Results}

\subsection{Basic molecules:  H$_2$, CO, and H$_2$O}
\begin{figure}
      \resizebox{\hsize}{!}{\includegraphics{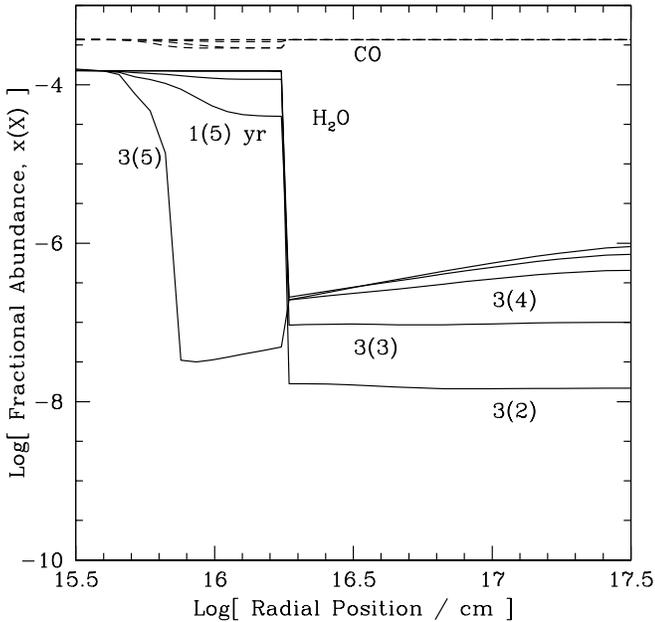}}
      \caption[]{The fractional abundances of CO and H$_2$O 
                 throughout the envelope as a function of time.
                 The dashed-lines correspond to the (constant)
                 CO abundance, and the solid lines to the
                 H$_2$O abundance.  The
                 curves are labeled by the time in years, where
		 $a(b) = a \times 10^{b}$.
              }
         \label{fig.xh2oxco.vsr}
   \end{figure}

Due to their stability, CO and H$_{2}$O are significant chemical
sinks, with abundances that are relatively constant with
time.  To see this, in Fig. \ref{fig.xh2oxco.vsr} we plot
the fractional abundance of CO and H$_2$O throughout the
envelope as functions of time.
As can be seen, the CO abundance is essentially constant in time.
The abundance has been chosen to be consistent with 
observations.

%
%

The water abundance in the warm interior is nearly constant,
due to the fact that the majority of the oxygen not in CO 
is initially placed into water.  This is consistent with models
we and others (e.g., Doty \& Neufeld \cite{DN},
Charnley \cite{C97}) have run which
show that even when the oxygen is not initially bound in water,
nearly all of the available oxygen is converted into 
water on a timescale of about one hundred years
due to fast neutral-neutral reactions in the warm gas.

The near discontinuity in the water abundance
at $T \sim 100$ K is due to the release of 
water from grain mantles.   This discontinuity
is consistent with observations of warm ($T \sim 300 - 500$ K) 
water in absorption toward AFGL 2591 
(Helmich et al. \cite{helmichetal1996}, Boonman
et al. \cite{boonmanetal2000}), with the lack of strong emission by cold
water at long wavelengths (Boonman et al. \cite{boonmanetal2000}), and by 
detailed modeling of the line emisson to be discussed in a
forthcoming paper (Boonman et al. -- in preparation).  

As noted by Charnley (\cite{C97}), the ion fraction and electron density
grow with time.  As seen in Fig. \ref{fig.xh2oxco.vsr} this
leads to a destruction of water on timescales of $> 10^{5}$ years
in the interior, in agreement with the results of Charnley (\cite{C97}).  
While the cosmic-ray ionization continually creates ions which
destroy water, reformation is temperature dependent.  A simple
extrapolation of the ``critical temperature'' for water formation
from Charnley (\cite{C97}) for our adopted cosmic-ray ionization
rate and density yields $180-200$ K.  Based upon the temperature structure
in Fig. \ref{tempdens}, this implies destruction of water for 
$r \sim 6-8 \times 10^{15}$ cm, in agreement with the results 
in Fig. \ref{fig.xh2oxco.vsr}.  It should be noted that the
destruction of water for $t > 10^{5}$ yrs is probably unimportant
for AFGL 2591 based both upon the water distribution inferred 
by Boonman et al. (in preparation), and the chemical evolution timescale
of $< 10^{5}$ yrs discussed in Sect. 5.

Finally, the growth in the water abundance with time in the exterior
occurs through slower (due to the lower abundances) ion-molecule
reactions.  Again, the ion-molecule reactions
are driven by cosmic-ray ionization.
In the exterior, average abundances of $< 3 \times 10^{-7}$ 
are achieved for $t \sim 3 \times 10^{4}$ years.

The results in Fig. \ref{fig.xh2oxco.vsr} have interesting
implications for the interpretation of water abundances. First, a simple
estimate of the water abundance inferred from the model radial column 
densities [assuming $x(\mathrm{H}_{2}\mathrm{O}) =
N(\mathrm{H}_{2}\mathrm{O})/N(\mathrm{H}_{2})$] would 
suggest a fractional abundance of water in our model of  
$x(\mathrm{H}_{2}\mathrm{O}) \sim 3 \times 10^{-5}$.  This
is a factor of 5 lower than the actual
water abundance adopted in the interior, and would by itself imply
a significantly different structure and chemistry involved.
This underscores the potential pitfalls in interpreting
column densities, as well as the importance of modeling the complete
physical, thermal, and chemical structure of the envelope 
in order to properly compare the relevant regions with observations.

A second implication is that beam
dilution can have an important effect on the inferred column densities.  
A simulated beam-averaged
column density commensurate with the beam of the
Submillimeter Wave Astronomy Satellite (SWAS) would imply 
a water abundance of 
$x(\mathrm{H}_{2}\mathrm{O}) = N(\mathrm{H}_{2}\mathrm{O})/
N(\mathrm{H}_{2}) = 10^{-7} - 10^{-8}$ depending upon the
time considered.  This low abundance is due to significant
beam-dilution from the small region of enhanced H$_{2}$O 
in the large beam.
The range of abundances is similar to that 
inferred by SWAS (see e.g., Snell et al. \cite{snelletalswas2000};
Melnick et al. \cite{melnicketalswas2000}; 
Neufeld et al. \cite{neufeldetalswas2000}).  Clearly, 
such an observation alone does not constrain the entire 
envelope. While it implies that a portion 
of the envelope (e.g., $T \le 100$ K)
has a low water abundance, 
it does not restrict the potential for a compact region 
of significant water abundance.

%
   \begin{figure}
      \resizebox{\hsize}{!}{\includegraphics{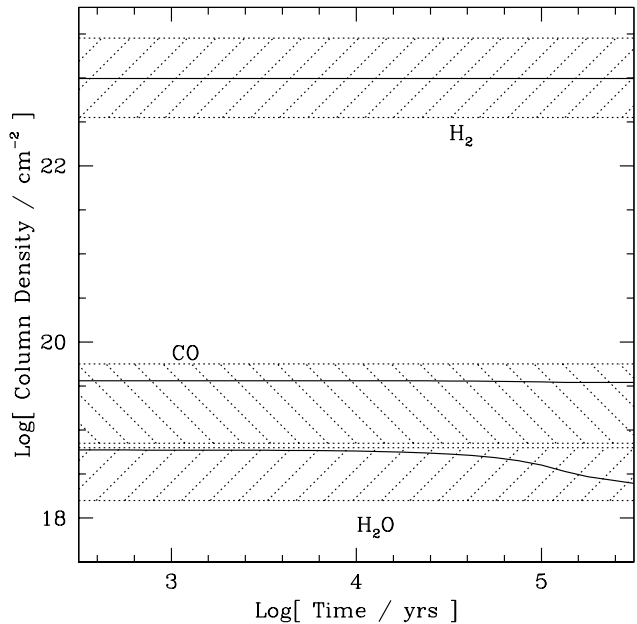}}
      \caption[]{The radial column densities of H$_{2}$, CO, and
                 H$_{2}$O as function of time (solid lines).
                 The shaded regions correspond to the observed 
                 abundances (with factor of two errorbars).
              }
         \label{fig.nh2nconh2o.vst}
   \end{figure}
%

As a comparison of the  column
densities with observations, in Fig. \ref{fig.nh2nconh2o.vst}
we plot the H$_{2}$, CO, and H$_{2}$O
column densities as a function of time.  
We assign errorbars of a factor of two consistent with the intrinsic 
uncertainties in the H$_{2}$O and CO results, and with the
fact that the H$_{2}$ results are scaled from the CO data,
as well as various radiative transfer effects.  
The ranges of the observed column densities are given by the
shaded regions.  As expected, our data match
the observed column densities within the uncertainties.

\subsection{Hydrocarbon and nitrogen chemistry}
Observations by Lahuis \& van Dishoeck
(\cite{Lv}) suggest that 
the $14 \mu$m bands of 
C$_{2}$H$_{2}$ and HCN 
are good tracers of hot gas.  
Perhaps more importantly,
the increase in observed column densities for
temperatures above a few hundred K implies that 
their chemistry may be altered at
high temperatures.
Since all of their inferred excitation
temperatures are well above the expected desorption
temperature of $\sim 100$ K, it is expected that
these enhanced abundances are due to warm gas-phase
chemistry.

\begin{figure}
    \resizebox{\hsize}{!}{\includegraphics{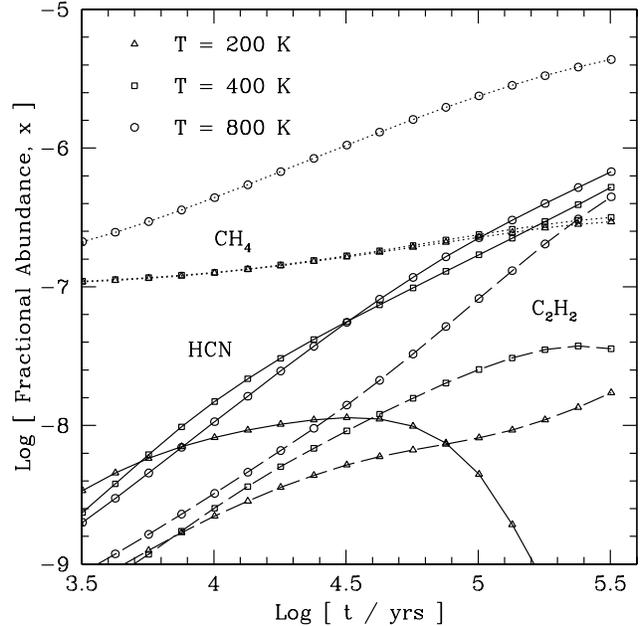}}
      \caption[]{The fractional abundances of HCN, CH$_{4}$, and 
                 C$_{2}$H$_{2}$ as functions of time for
                 various temperatures.  Here $n(\mathrm{H}_{2})=10^{7}$
                 cm$^{-3}$.  Notice the general enhancement of
                 the abundances with increasing temperature.
              }
         \label{fig.xhcnxch4xc2h2.vst.hotmodels}
   \end{figure}

In order to test this, we have constructed single-position
models for the chemistry at $n(\mathrm{H}_{2})=10^{7}$ cm$^{-3}$
and $T \geq 200$ K.  The results are shown 
in Fig. \ref{fig.xhcnxch4xc2h2.vst.hotmodels}.  Clearly, 
higher temperatures do increase the abundances of simple
hydrocarbons and nitrogen-bearing species, with higher
abundances prevalent once $T \sim$ few hundred K.

The enhanced HCN abundance is similar to that found by
Rodgers \& Charnley (\cite{RC}).  In parallel with
their work, we find that the 756 K endothermic 
reaction CN + H$_{2} \rightarrow$
HCN proceeds quickly for $T>200$ K, producing significant HCN.
However, while Rodgers \& Charnley (\cite{RC}) assume the reaction
C$^{+} + $NH$_{3}$ favors HCNH$^{+}$ (following
ab initio calculations by Talbi \& Herbst \cite{th}), we
assume that H$_{2}$NC$^{+}$ is the favored product to 
account for the observed HNC/HCN abundance ratio in many sources.
In our case, then, the CN is formed via the neutral-neutral reaction 
$\mathrm{CS}+\mathrm{N}\rightarrow\mathrm{CN}+\mathrm{S}$.
This reaction has a barrier of $1160$ K, leading to
significant production for temperatures above 200 K.  Overcoming
these barriers can increase the abundance from a peak
of $10^{-8}$ at 200 K, to $\sim 10^{-7}$ for 
$t > 4 \times 10^{4}$ years, and $\sim 10^{-6}$ for 
$t > 3 \times 10^{5}$ years for $T \geq 400\mathrm{K}$. 

%
%

Methane shows perhaps the most dramatic increase in 
abundance at very high temperatures.  In fact, 
methane contains more carbon at the latest times 
than all species other than CO$_{2}$ and CO 
initially.  This is due to the fact that ion-molecule 
reactions driven by cosmic-ray ionization (e.g., 
$\mathrm{He}^{+}+\mathrm{CO}\rightarrow\mathrm{C}^{+}+\mathrm{O}$)
can produce C$^{+}$.  This then reacts via carbon insertion
(Herbst \cite{herbstreview1995}) 
with H$_{2}$ to form CH$^{+}$ at high temperatures, and then
in a chain with H$_{2}$ up to CH$_{3}^{+}$, 
which dissociatively recombines to form CH.  While
CH$_{3}^{+}$ can also dissociatively recombine to form CH$_{2}$, 
the dominant pathway to CH$_{2}$ at high temperatures is 
$\mathrm{CH}+\mathrm{H}_{2}+1760\mathrm{K}
\rightarrow\mathrm{CH}_{2}+\mathrm{H}$.   
Reactions with H$_{2}$ then produce CH$_{4}$
(overcoming barriers of 6400 K and 4740 K to form 
CH$_{3}$ and CH$_{4}$ respectively),  
leading to abundances of $1-3 \times 10^{-7}$ for
$T \leq 400\mathrm{K}$.  
However, once the temperature
increases to $\sim 600-800 \mathrm{K}$, abundances
can reach $x(\mathrm{CH}_{4}) \sim 10^{-6}$ at 
$t \sim 3 \times 10^{4}$ years. 

Acetylene is also enhanced at high temperatures.
The pathway here is similar to that in diffuse
and dark clouds (van Dishoeck \& Hogerheijde \cite{vH}).  
However, in our model, 
acetylene is formed via reactions of water
with C$_{2}$H$_{3}^{+}$ instead of dissociative recombination.  
A second difference
is that C$_{2}$H$_{3}^{+}$ 
is produced via $\mathrm{C}_{2}\mathrm{H}_{4}+\mathrm{H}_{3}^{+}$.
While the ``usual'' $\mathrm{CH}_{4}+\mathrm{C}^{+}$
production route still occurs, the destruction of 
C$_{2}$H$_{4}$ by O is reduced as the temperature 
increases due to the fact that the oxygen is quickly 
converted into water by neutral-neutral reactions (see Sect.\ 4.1).
Again, cosmic-ray ionization, carbon insertion, and water play a role, 
both in the production of H$_{3}^{+}$, and in the
production of C$_{2}$H$_{4}$ via
$\mathrm{CO}\rightarrow\mathrm{C}^{+}\rightarrow \ldots
\rightarrow \mathrm{CH}_{3}^{+} + \mathrm{CH}_{4}
\rightarrow \mathrm{C}_{2}\mathrm{H}_{5}^{+} + 
\mathrm{H}_{2}\mathrm{O} \rightarrow 
\mathrm{C}_{2}\mathrm{H}_{4}$.  The enhanced C$_{2}$H$_{2}$
abundance is in the range 
$5 \times 10^{-9} \leq x(\mathrm{C}_{2}\mathrm{H}_{2})
\leq 2 \times 10^{-8}$ for $200 \mathrm{K} \leq T \leq 800 \mathrm{K}$
at $t = 3 \times 10^{4}$ years.  At late times, it is
almost always less than $3 \times 10^{-8}$ 
at 200 K, less than $5 \times 10^{-8}$ at 400 K, and can
reach $5 \times 10^{-7}$ at 800 K. 

\begin{figure}
    \resizebox{\hsize}{!}{\includegraphics{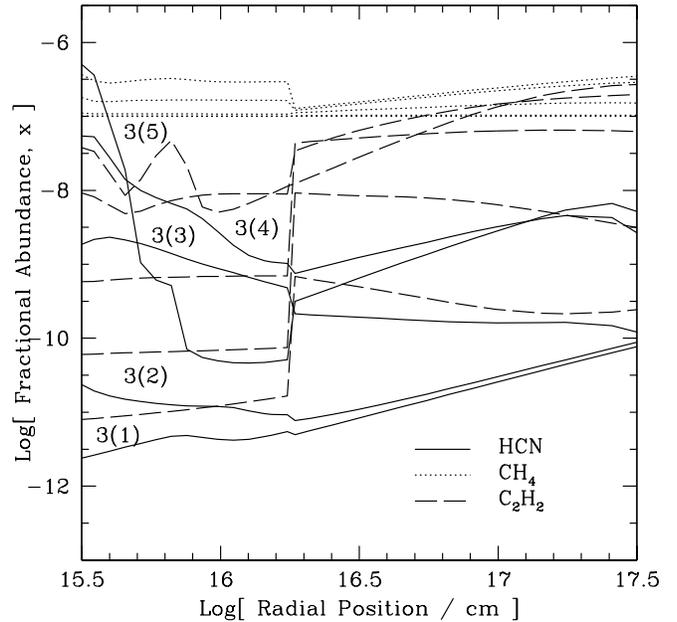}}
      \caption[]{The time evolution of the fractional abundances of 
                 HCN, CH$_{4}$, and C$_{2}$H$_{2}$ throughout
                 our model, incorporating the temperature and
                 density distributions desribed in the text.
                 The HCN data are labeled by the time in years,
                 where $a(b) = a \times 10^{b}$.  The times for
		 each curve increase upward at the inner radial position.
                 Note the enhancement of the abundances in the
                 warmer interior.
              }
         \label{fig.xhcnxch4xc2h2.vsr}
   \end{figure}

In order to see how this high-temperature chemistry pertains
to our model, in Fig. \ref{fig.xhcnxch4xc2h2.vsr} we plot
the fractional abundances of HCN, CH$_{4}$, and 
C$_{2}$H$_{2}$ throughout the envelope for various times.  
As expected from the previous discussion, we see
enhanced abundances of HCN, C$_{2}$H$_{2}$, and CH$_{4}$, 
especially in the warm interior.  
The enhancement of C$_{2}$H$_{2}$ in the exterior  
has two primary causes. 
First, in this region C$_{2}$H$_{2}$ is primarily formed
via dissociative recombination of C$_{2}$H$_{3}^{+}$. 
The destruction of C$_{2}$H$_{3}^{+}$ by O has a 215 K barrier 
that cannot be
overcome in the cool exterior, leaving more  C$_{2}$H$_{3}^{+}$
to produce acetylene. 
Second, an alternate production pathway via  
C$_{3}$H$_{3}^{+} + $O$ \rightarrow $C$_{2}$H$_{2}$ is
enhanced in the exterior due to our increased initial O 
abundance in that region (see Table \ref{modelparameters}).

Cosmic-ray driven ion-molecule
chemistry again plays a role for $t > 10^{5}$ years. 
In particular,
the destruction of HCN near $10^{16}$ cm is due to reactions with
HCO$^{+}$.  For C$_{2}$H$_{2}$ both HCO$^{+}$ and O are important
destruction reactants near $10^{16}$ cm.  The enhancement in 
C$_{2}$H$_{2}$ near $r \sim 6-8 \times 10^{15}$ cm is due
to a decrease in atomic oxygen at this position for late times
(see also Sect. 4.6).
 
Observations of high-lying HCN lines in the submillimeter were undertaken
by Boonman et al. (\cite{boonmanetal2001}).  They utilized a sophisticated 
radiative transfer model of the excitation, line shapes
and strengths to analyze their data,
and suggested that HCN follow a ``jump'' model,
with an abundance of $x(\mathrm{HCN}) \sim  1 \times 10^{-6}$ for
$T \gos 230 \mathrm{K}$, and $x(\mathrm{HCN}) \sim 10^{-8}$ for
the cool exterior.  The results in 
Figs. \ref{fig.xhcnxch4xc2h2.vst.hotmodels} and
\ref{fig.xhcnxch4xc2h2.vsr}
are consistent with this supposition, with abundances of 
a  few $\times 10^{-7}$ at high temperatures, and $\sim 10^{-8}$ 
at lower temperatures 
and later times.

As expected, the results in Fig. \ref{fig.xhcnxch4xc2h2.vsr}
are not as dramatic as in 
Fig. \ref{fig.xhcnxch4xc2h2.vst.hotmodels}, as our
physical and thermal model only extends into $T \sim 440\ \mathrm{K}$,
less than the temperatures at which the greatest enhancements
occur.  
Consequently, care must be used when comparing the
results with observations, 
as the different temperature
components may not necessarily probe the portions of the region
being modeled.

%
%

%
   \begin{figure}
    \resizebox{\hsize}{!}{\includegraphics{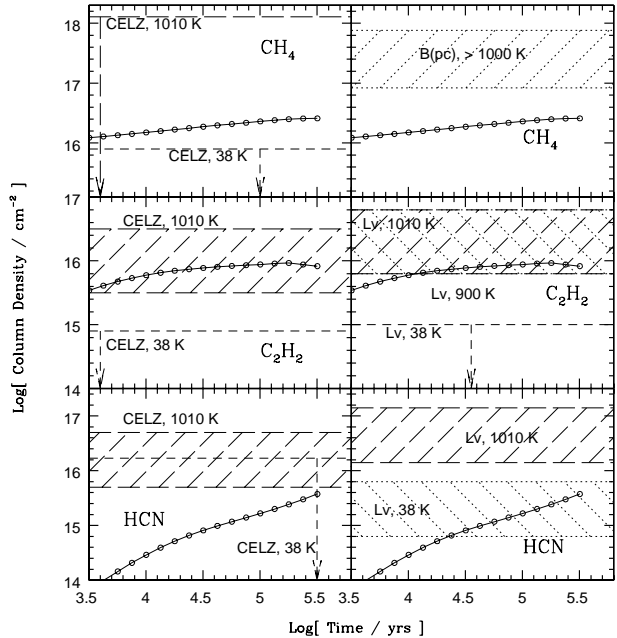}}
      \caption[]{A comparison of the predicted and observed
                 column densities of HCN, CH$_{4}$, and C$_{2}$H$_{2}$ 
                 as a function of time.  The model predictions
                 are given by the solid lines and accompanied by
                 the filled circles.  The observations are divided 
		 into two groups.  The left-hand panels are for
                 the infrared data of Carr et al. (\cite{CELZ} CELZ), 
		 while the right-hand panels
                 are for ISO data from Lahuis \& van Dishoeck
		 (\cite{Lv} Lv) and Boonman (B(pc), private communication).  
		 Data which are upper limits are signified by 
		 downward arrows.
                 Other data have been given an arbitrary factor
                 of 3 uncertainty, and are given by the shaded regions.
              }
         \label{fig.nhcnnch4nc2h2.vst}
   \end{figure}
%

Such a comparison is given in Fig. \ref{fig.nhcnnch4nc2h2.vst}.
Here the model predictions for CH$_{4}$, C$_{2}$H$_{2}$, and 
HCN are compared with the infrared observational data, which probe
column density.  In the left-hand
panels we compare to the data of Carr et al. (\cite{CELZ}),
omitting the 200 K data as these arise in the outflow.
In the right-hand panels we compare to the data
of Boonman (private communication),
and Lahuis \& van Dishoeck (\cite{Lv}).
  
When we compare with the lower temperature data,
the CH$_{4}$ model results are close to  
the observed error bounds.  
On the other hand, they are well above the high 
temperature component of the observations.  
This is not a suprise, as the CH$_{4}$
chemistry is relatively unaffected below about 400 K, with 
very significant production at higher temperatures.

On the other hand, 
the C$_{2}$H$_{2}$ data fit the high temperature components
of the observations.
This seems to imply that while high temperature chemistry
can be important, the effects are 
noticably smaller than for CH$_{4}$, consistent with
the results of Fig. \ref{fig.xhcnxch4xc2h2.vsr}.  In particular,
the fact that the predicted column density is so much
higher than the low-temperature column density suggests
that warm chemistry can enhance C$_{2}$H$_{2}$, while
the fact that the predicted column densities fall in the
lower range of the observed values suggests that there
exists room for some enhancement ($\sim 3-5 \times$) 
in the C$_{2}$H$_{2}$ abundance at higher temperatures, 
consistent with Fig. \ref{fig.xhcnxch4xc2h2.vsr}. 

Finally, in the lower panels of Fig. \ref{fig.nhcnnch4nc2h2.vst}
we show the comparison for HCN.  
Here we see the potential
for further importance of high-temperature chemistry.
In the lower-left panel, the HCN
model prediction is consistent with the upper limit derived by
Carr et al. (\cite{CELZ}) at low temperatures.   
Similarly, in the lower-right panel, the predicted column densities
are consistent with the low-temperature component fit by
Lahuis \& van Dishoeck (\cite{Lv}).  In both cases, it appears
that our model reproduces the production of cool HCN 
quite well.

On the other hand, the model predictions are well below
the observed column densities for the hot components in 
each of the panels.
This is most probably due to 
the significant production of HCN at temperatures above
400 K (see above).  
This is further supported by the fact that when a single
temperature component is determined by Lahuis \& van
Dishoeck (\cite{Lv}), they find 
$T \sim 600 \ \mathrm{K}$.  Taken at face value, their data suggest that  our
model does not extend inward far enough to include this
hot gas.  

At this point, one may ask if a simple extension of
our power-law model inward would 
increase the temperature and column density sufficiently to 
fit the observed HCN data 
(i.e., at 1010K).  We have examined this possibility
by extending our model inward, with no success.  
While a fractional abundance of $x(\mathrm{HCN}) \sim 10^{-7}$
would reproduce the data, the conditions necessary would also 
produce a water column density 
$N(\mathrm{H}_{2}\mathrm{O}) \sim 4 \times 10^{19}$
cm$^{-2}$, over an order of magnitude above the observations.

An alternative solution is to adopt a ``flattened'' 
(i.e., $n(r) \propto r^{0}$) 
density profile for $r<r_{\mathrm{in}}$.  In this
case, the extra column of water would be consistent with the
observations, and the column of HCN would vary as 
$N(\mathrm{HCN}) \sim 2 \times 10^{15}(x(\mathrm{HCN})/10^{-7})$.  
While the column
could be fit if $x(\mathrm{HCN}) = 10^{-6}$, this is inconsistent
with the results of Fig. \ref{fig.xhcnxch4xc2h2.vsr}.  First, the
chemistry does not show strong variation between 400 and 800~K,
suggesting that high temperatures alone will not produce significantly
more HCN.  Furthermore, to achieve $x(\mathrm{HCN}) = 10^{-6}$ at
these temperatures would require an extended time for chemical
evolution in the interior, and would be inconsistent with the
abundances of other observed species (see Sect. 5).

There are four possible resolutions to this difficulty.  First, 
and least likely, is the possiblity that the chemical evolution
time in the interior is somehow longer than in the exterior.  
We can think of no way in which this may occur. 
The second possibility is that the hydrocarbon and 
nitrogen chemistry is currently incomplete, especially at
high temperatures.  If another pathway to producing HCN
exists above about 600 K, it would be possible to have
abundances of $10^{-6}$. 
Third, there is the possibility that HCN is present in grain
mantles, and is injected into the hot gas.  Though this
is expected to be unimportant (van der Tak et al. \cite{vdt1999}),
it may conceivably play a small role.

The fourth, and perhaps most likely, possibility is that
there exists some as of yet unidentified destruction
mechanism for water at high temperatures.  This would 
remove the problem of the overly-large water column if
the envelope were to simply extend further inward.  
It is possible that evidence exists for this.  
As discussed by van Dishoeck (\cite{vfaraday})
observations of water gas and ice toward various sources show
significantly less total water in  
hotter sources than in cooler sources.  Given 
our current understanding of the chemistry of H$_{2}$O 
production, it would be easiest to explain this effect if
there existed a mechanism for H$_{2}$O destruction
at high temperatures.
Further study into the high temperature chemistry of 
water, hydrocarbons, and nitrogen-bearing species
would be of significant
importance in understanding this problem.

\subsection{Sulphur chemistry}
The chemistry of sulphur in hot cores 
is well-described by Charnley (\cite{C97}). 
In our model, we have adopted a chemistry and
set of initial conditions (in the warm region) which is
similar.  However, given the fact that his model 
was for a single point in space, while our model extends over
a range of physical and thermal parameters, and given recent
observations of sulphur-bearing molecules toward AFGL 2591, we
present our results here.

%
   \begin{figure}
    \resizebox{\hsize}{!}{\includegraphics{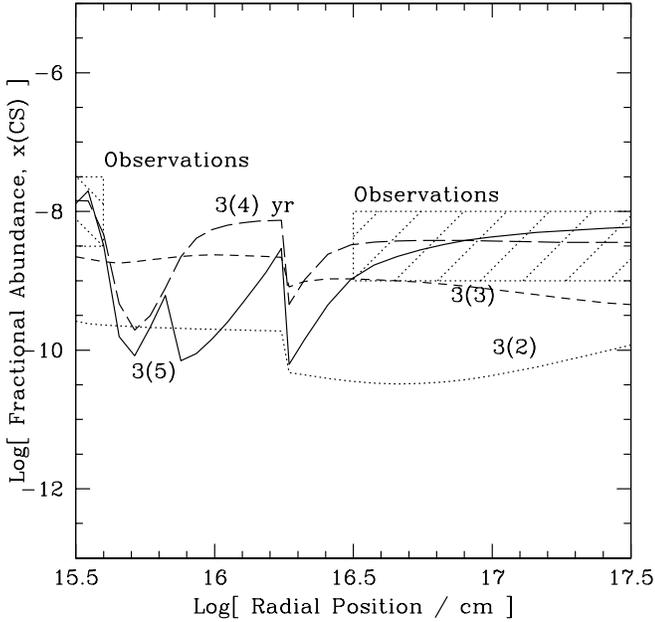}}
      \caption[]{The fractional abundance of CS throughout
                 the envelope for various times.  Nearly half
                 of the sulphur is in CS at late times in the
                 cool exterior, essentially ``fixing'' the 
                 gas-phase sulphur abundance.  The agreement
                 with observations in the warm interior, 
                 however is not fixed.  The curves are labeled
                 by the time in years, where $a(b) = a \times 10^{b}$.
              }
         \label{fig.xcs.vsr}
   \end{figure}

In the cool exterior of our model we find that there exist
a large number of pathways to shuttle sulphur into CS.  The
end product is that approximately 50\% of the sulphur is
transformed into CS by $t \sim 10^{5}$ years.  This is shown 
in Fig. \ref{fig.xcs.vsr}, where we plot the fractional
abundance of CS for various times.  
No single production reaction accounts for more than 
25\% of the final CS abundance.  This means that, at late
times at least, CS is a good measure of the sulphur 
abundance in the exterior.  To accomodate this fact, and in
order to match observations of the CS abundance (see
Table \ref{molecularobs}), we adjust the initial 
sulphur abundance to $x(\mathrm{S}) = 6 \times 10^{-9}$ for
$T \leq 100 ~ \mathrm{K}$.  This produces a nearly constant
abundance in the exterior in good agreement with the observations.

In the interior, the CS abundance increases at 
intermediate and 
late times to
$x(\mathrm{CS}) \sim 10^{-8}$.  This is also in agreement 
with the observations.  However, while the abundance in the
exterior is essentially ``forced'' by our initial sulphur 
abundance, the fraction in CS in the interior is not.

The variation in the interior CS abundance
(as with water and the hydrocarbons)
is again related to the oxygen and cosmic-ray
driven ion-molecule chemistry.  
In particular the dip near 
$r \sim 5 \times 10^{16}$ cm is due to the increased atomic
oxygen abundance (see Sect. 4.6) in this region.  This leads
to more conversion of sulphur to SO$_{2}$, and thus less
to CS.  The decrease in CS abundance near $10^{16}$ cm
at $t = 1-3 \times 10^{5}$ yrs is due to the fact that there 
is less OH available for conversion of sulphur out of H$_{2}$S to CS.

%
   \begin{figure}
    \resizebox{\hsize}{!}{\includegraphics{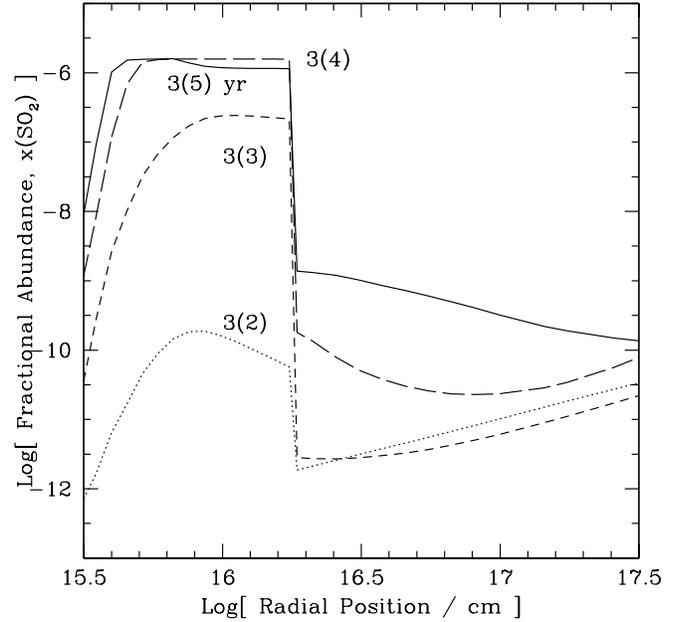}}
      \caption[]{The fractional abundance of SO$_{2}$ throughout
                 the envelope at various times.  Note the increase
                 near $T = 100\ \mathrm{K}$ as the free sulphur
                 is forced into SO$_{2}$.  The decrease at
                 high temperatures is due to the loss of the
                 reactant OH via its more efficient inclusion
                 into water at those temperatures.
                 The curves are labeled by the time in years, 
                 where $a(b) = a \times 10^{b}$.
              }
         \label{fig.xso2.vsr}
   \end{figure}

The sulphur abundance in the warm ($T \sim 100 - 400 ~ \mathrm{K}$) gas
is well-determined by the SO$_{2}$ abundance.  In our model,
SO$_{2}$ is formed by 
$\mathrm{H}_{2}\mathrm{S} + \mathrm{(OH, H)} \rightarrow 
\mathrm{HS} + \mathrm{O} \rightarrow 
\mathrm{SO} + \mathrm{OH} \rightarrow \mathrm{SO}_{2}$.
The initial reactions of H$_{2}$S with H and OH have
barriers of 352 K and 80 K, respectively. 
As a result, little SO$_{2}$ is produced in the cool exterior,
while the barriers can be overcome in the interior leading to 
significant SO$_{2}$ production. 
As the temperature further increases, however, the 
OH can be more easily forced into water, leaving little for the
$\mathrm{SO}+\mathrm{OH} \rightarrow \mathrm{SO}_{2}+\mathrm{H}$
reaction.  
This can be seen in the very interior of 
Fig. \ref{fig.xso2.vsr}, where the SO$_{2}$ abundance drops
at high temperatures.  In our model approximately 90\% of the 
sulphur returns to atomic form at $\sim 440 \ \mathrm{K}$, with 
approximately 10\% in H$_{2}$CS, and a few percent in CS and 
OCS.

While we are are unable to identify the sulphur resevoir 
assuming solar abundances roughly hold, it appears
that a significant portion would need to exist in or on dust grains.
Under this constraint, we can also identify
SO$_{2}$ as the primary sink of molecular sulphur in warm
(100-300 K) gas (assuming no O$_{2}$ is released during
heating of the grain mantles -- Charnley \cite{C97}).  
As a result, the sulphur abundance in warm
molecular gas at later times can be approximately determined by the 
SO$_{2}$ abundance.  In our model, this requires the adjustment
of the initial H$_{2}$S abundance from the value of 
$10^{-7}$ adopted by Charnley (\cite{C97}) to $1.6 \times 10^{-6}$.
This value is, coincidentally, similar to the H$_2$S gas-phase
abundance seen by
Minh et al.\ (\cite{minhetal1990}) toward Orion.  A comparison of our model
predictions with observations by Keane et al. (\cite{keaneetal})  
show similar column densities of 
$4 \times 10^{16}$ cm$^{-2}$ and $6 \times 10^{16}$ cm$^{-2}$
respectively.  
It is also intruiging that the  
excitation temperature inferred by 
Keane et al. (\cite{keaneetal})
for SO$_{2}$ toward
AFGL 2591 is $\sim 750\mathrm{K}$, suggesting formation 
in a warm dense region of a few hundred K.

\subsection{CO$_{2}$ chemistry:  potential heating events?}
%
   \begin{figure}
    \resizebox{\hsize}{!}{\includegraphics{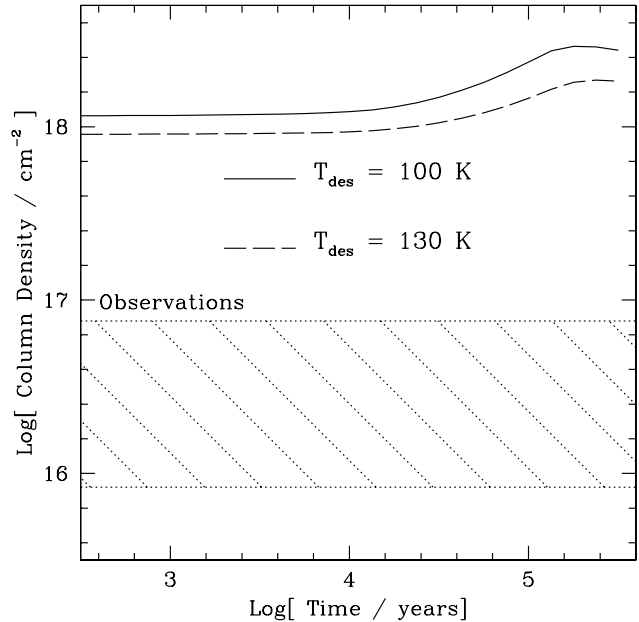}}
      \caption[]{The column density of CO$_{2}$ predicted by 
                 our base model as a function of time (solid  
                 and dashed lines) for two different assumed
		 desorption temperatures.  No impulsive heating
		 event is assumed (see text).  The observed column 
                is shown by the shaded region.  Notice
                the extent to which the model overpredicts
                 CO$_{2}$.
              }
         \label{fig.nco2.vst}
   \end{figure}
%
An important problem in the chemistry of the envelopes
of massive young stars is the low observed gas-phase abundance
of CO$_{2}$ (see e.g., van Dishoeck \& van der Tak \cite{vvkorea}).
Observations by ISO indicate large solid CO$_{2}$
abundances (de Graauw et al. \cite{deGetal}, 
Whittet et al. \cite{whittetelias16},
Ehrenfreund et al. \cite{ehrenfreundetal}, 
Gerakines et al. \cite{gerakines}), 
with a CO$_{2}$/H$_{2}$O 
abundance in the ice mantles of $10-20\%$.  In the
warm regions close to the protostars, these mantles should
be evaporated.  Assuming water ice abundances of
a few $\times 10^{-5}$ (Tielens et al. \cite{tielens91}; 
Gensheimer et al. \cite{gensheimer})
implies a liberated fractional abundance of 
$x(\mathrm{CO}_{2}) \sim 10^{-5} - 10^{-6}$.  On the other
hand, ISO observations of gas-phase CO$_{2}$
(van Dishoeck et al. \cite{vdetal96}, 
Boonman et al. \cite{boonmanetal2000})
suggest $x(\mathrm{CO}_{2}) \sim 10^{-7}$.  These
results indicate that CO$_{2}$ is quickly destroyed
after evaporation from ice mantles.

To see this discrepancy between the amount of CO$_{2}$ predicted
in our base model and that observed, in Fig. \ref{fig.nco2.vst} we plot
the predicted CO$_{2}$ column density as a function of
time.  Also plotted are the observations of Boonman
et al. (\cite{boonmanetal2000}).  Clearly, the base model significantly
overpredicts the CO$_{2}$ column density in AFGL 2591,
confirming the general results above.

Charnley \& Kaufman (\cite{CK}) studied destruction of
CO$_{2}$ by both H and H$_{2}$, suggesting that 
destruction of CO$_{2}$ by H in postshock flows could be important.
In order to test this, we have constructed models with 
non-zero atomic hydrogen abundances, as would be expected in 
partially dissociative
shocks.  While the CO$_{2}$ can be effectively destroyed on 
a shock cooling timescale of $\sim 30$ yrs, CH$_{4}$,
NH$_{3}$, and H$_{2}$O can be destroyed even more efficiently.
While this does not pose a significant problem for CO$_{2}$
or NH$_{3}$ which have low observed abundances or upper limits,
there are effects on other species.  
In particular, 
in dissociative shocks
the water column density is decreased by a factor of 2-3.
Furthermore, once destroyed, only little water is re-formed in the range 
$100 \leq T(\mathrm{K}) \leq 300$, inconsistent with the
results of Boonman et al. (in preparation).
Similarly, the O and O$_{2}$ abundances are significantly 
increased.  
On the other hand, 
the CH$_{4}$ abundance is decreased by an order
of magnitude in the interior.  
This process only requires a few percent H$_{2}$ dissociation.

A second potential difficulty is that it is unclear if 
a large enough fraction of the envelope can be 
disturbed by a shock to significantly affect the 
global CO$_{2}$ abundance, as evidenced by the
relatively small line-widths in much of the envelope 
(van der Tak et al. \cite{vdt1999}).

Doty et al. (\cite{dotyco2}) reconsidered
this problem in light of previously unused laboratory measurements of
the destruction of CO$_{2}$ by H$_{2}$ 
(Graven \& Long \cite{gravenlong}).  They
found that destruction by H$_{2}$ may dominate destruction by H in the
very warm gas, near T $\sim 1000-1600 ~ \mathrm{K}$.  While this may
occur in a number of ways, Doty et al. (\cite{dotyco2}) considered two
possibilities: a uniform temperature increase (such as from the
passage of a $v \sim 20 - 30$ km s$^{-1}$ MHD shock -- Draine
et. al. \cite{draineetal}), and a central luminosity increase caused,
for example, by an accretion (FU-Orionis-type) event.  The
possibility of impulsive heating events may be supported by evidence from
continuum emission by crystalline silictes (Smith et
al. \cite{smithetal}, \& Aitken et al. \cite{aitkenetal}) which
suggests that an annealing event may have occurred in AFGL 2591.  If
such a heating event occurred, Doty et al. (\cite{dotyco2}) find that
it is possible for the CO$_{2}$ to be removed on a timescale of
$10^{0}-10^{4}$ years by H$_2$.  Recent calculations of the potential
surface for the CO$_2$ + H$_2$ reaction suggest, however, that the
barrier for the reaction may be higher than indicated by the old
laboratory experiments, so that this issue remains unsettled (Talbi \&
Herbst \cite{talbiherbst2002}).  Clearly, 
further laboratory studies of this reaction at high
temperatures are urgently needed.

While speculative, destruction of CO$_{2}$ by H$_{2}$ in
this fashion has some advantages.  First, 
there is very little atomic hydrogen available to 
affect the chemistry, and in particular to influence 
CH$_{4}$, O, O$_{2}$, and H$_{2}$O.  
Second, and perhaps more importantly, variations in the observed
column density of CO$_{2}$ may potentially be explained
by variations in the size and/or duration of the 
proposed heating event -- depending upon its origin, or
the time since the heating event and the local cosmic-ray
ionization rate.

As a final note, it is interesting to also consider the possibility
that the CO$_2$ desorption temperature may be greater than 100 K.
Recent work by Fraser et al.  (\cite{fraseretal2001}), suggests that
the desorption temperature of water may be as high as $120-130$ K.  If
the solid CO$_{2}$ is contained in a water-ice matrix as suggested by
observations (Gerakines et al. \cite{gerakines}), then it may be
interesting to consider the effect of this higher desorption
temperature on $N(\mathrm{CO}_{2})$.  In Fig. \ref{fig.nco2.vst}, we
present predicted column densities for the re-formation of CO$_{2}$,
assuming desorption temperatures of both 100 K and 130 K.  The effect
is a decrease in the CO$_2$ column densities by a factor of two,
insufficient to explain the discrepancies.

\subsection{Cosmic-ray ionization rate}
As discussed earlier, cosmic-ray ionization can play
an important role in driving ion-molecule chemistry at later times.
In our model, we adopt the cosmic-ray ionization rate
for AFGL 2591 of $\zeta=5.6\times 10^{-17}$ s$^{-1}$
as determined by van der Tak \& van Dishoeck 
(\cite{vv}).
While the cosmic ray flux is unique, the ionization rate will vary
with position if the particles are absorbed.  As evidence for
cosmic ray absorption is inconclusive 
(see, e.g., van der Tak \cite{vdtboulder}), we adopt a single
cosmic ray ionization rate for AFGL 2591. 

%
   \begin{figure}
    \resizebox{\hsize}{!}{\includegraphics{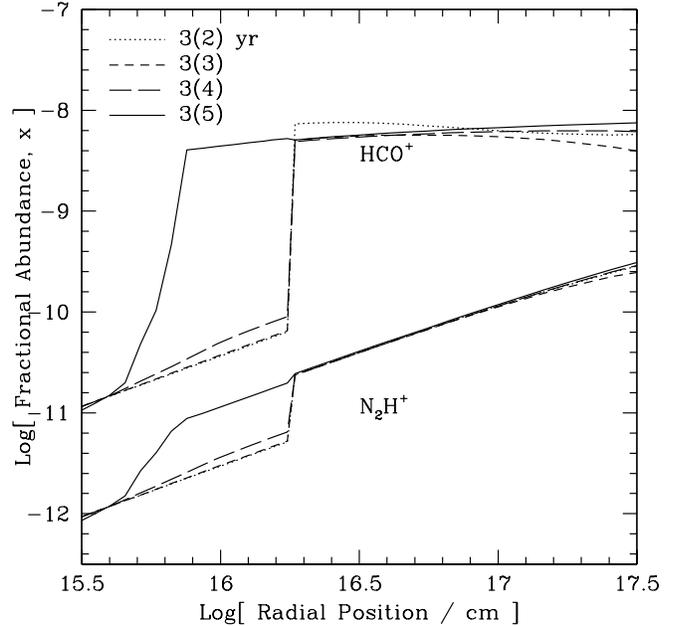}}
      \caption[]{The fractional abundance of HCO$^{+}$
                 and N$_{2}$H$^{+}$      
                 througout
                 the envelope for various times.  Note the
                 marked decrease about 100 K, where reactions
                 with H$_{2}$O become important.
                 The curves are labeled by the time in years,
                 where $a(b) = a \times 10^{b}$.
		  
              }
         \label{fig.xhcopxn2hp.vsr}
   \end{figure}

In Fig. \ref{fig.xhcopxn2hp.vsr}
we plot the predicted fractional abundance of HCO$^{+}$ and
N$_{2}$H$^{+}$.  
There are two important features.
First, there is significant destruction of 
HCO$^{+}$ at the water desorption position, 
due to the reaction
$\mathrm{HCO}^{+}+\mathrm{H}_{2}\mathrm{O}\rightarrow
\mathrm{H}_{3}\mathrm{O}^{+}+\mathrm{CO}$.  This is in
agreement with the model of van der Tak \& van Dishoeck (\cite{vv}).
While they argue that this jump in abundances is not important
in constraining the cosmic ray ionization rate, our overall
HCO$^{+}$ abundance is consistent with their observations,
and thus lends support
to their somewhat high value for $\zeta$
in AFGL 2591.  
The situation is similar for N$_{2}$H$^{+}$.
Second, at $t = 3 \times 10^{5}$ yrs, the ion abundances
increase in the interior.  This is consistent with 
Charnley (\cite{C97}), and is due to the fact
that the cosmic-ray ionization continues to produce more ions, 
which eventually destroy a significant fraction of the complex
molecules up to the position where the temperature is high
enough to re-form them. 

The cosmic-ray ionization rate also affects the abundance
of H$_{3}^{+}$.  In our models, reasonable time ($t \geq 10^{3}$ years)
column densities are $\sim 2.6\times 10^{13}$ cm$^{-2}$, almost
an order of magnitude below those observed by McCall et al. 
(\cite{mccalletal}).
If a comparison of these results were used to infer a cosmic-ray
ionization rate, one would obtain a much larger value.
While large H$_{3}^{+}$ abundances in the diffuse ISM have
been reported by McCall et al. (\cite{mccalletal2002}) -- which 
they suggest may be due to uncertainties in dissociative
recombination rate -- van der Tak \& van Dishoeck (\cite{vv}) 
have also noted there exists a variation in H$_{3}^{+}$ column
density with distance which suggests that intervening clouds may be
important.

The cosmic-ray ionization rate
also affects the HCN abundance.
Decreasing $\zeta$ makes it harder to form HCN.  For example
lowering $\zeta$ by a factor of three decreases the enhancement
of HCN by a factor of three even at 800 K, placing the warm HCN
abundance at $\leq 3 \times 10^{-7}$ --
well below the observations.
Furthermore, the same change also increases the time for the cold
column of HCN to reach the observed range to $t \sim 10^5$ yrs, in
disagreement with the age constraints discussed below in 
Sect. 5.

Based upon these results, it appears
that the value of $\zeta = 5.6 \times 10^{-17}$ s$^{-1}$
inferred by van der Tak \& van Dishoeck (\cite{vv})
is correct to within a factor of three.
Any value much lower would significantly hamper the
production of HCN, making for disagreement with the
observations.  Any value much higher would 
be in conflict with the observed ion abundances.

\subsection{Other species}
\begin{figure}
    \resizebox{\hsize}{!}{\includegraphics{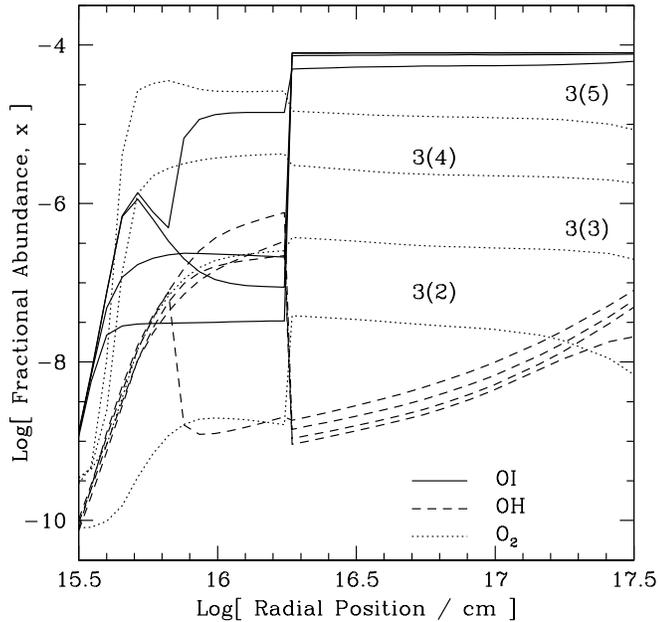}}
      \caption[]{The fractional abundance of O, OH, and O$_{2}$
            throughout the envelope for various times.  The
	    curves are labeled by the time in years, where
	    $a(b) = a \times 10^{b}$.  The times increase
	    upward.  The only exception is the dip for
	    OH near $10^{16}$ cm, which corresponds to 
	    $t=3 \times 10^{5}$ yrs.
              }
         \label{fig.xoixohxo2.vsr}
   \end{figure}

As oxygen and oxygen-bearing species can have a signifcant effect
on the chemistry, in Fig. \ref{fig.xoixohxo2.vsr} we 
plot the fractional abundances of O, OH, and O$_{2}$ as
functions of position for various times.  

%

The increase in the atomic oxygen abundance near 
$10^{16}$ cm is due to the fact that O is freed from
water at late times via ion-molecule reactions
with H$_{2}$O and CO as discussed in Sect. 4.1.  
As the water is destroyed, the main production 
mechanism for OH [(HCO$^{+}$, ~ H$_{3}^{+}$) $+$
H$_{2}$O $\rightarrow$ H$_{3}$O$^{+} + $e$^{-}
\rightarrow$ OH] is removed.  This leads to a
decreased OH abundance at this position at late times.

The peak in the atomic oxygen abundance near 
$r \sim 5-8 \times 10^{15}$ cm is due to the competition
between production of O by ion-molecule reactions with CO, and the
destruction of O at high temperatures by reactions with
OH and H$_{2}$.  Once the temperature reaches $\sim 180-200$ K,
neutral-neutral re-formation of water can balance
the destruction by ion-molecule reactions on these timescales,
as discussed in Sect. 4.1, and in Fig. \ref{fig.xh2oxco.vsr}.
This leads to a greater OH abundance at these positions,
and thus a decreased O abundance.  

In any case, the excess atomic oxygen is easily converted
to molecular oxygen over time at temperatures less than
300 K.  This places an important constraint on the
temporal evolution of the source as discussed in Sect. 5 below.

It is also interesting to note that the dominant nitrogen
resevoir is molecular nitrogen.  While atomic
nitrogen is somewhat abundant (see Table \ref{otherspecies}), only
about 1 \% or less of the nitrogen is in atomic form -- and
that preferentially at later times.   

Although we have endeavored to consider detailed 
comparisons between our model predictions and 
observations, a worthwhile test of any model is the
predictions it makes for future observations.
Consequently, in Table \ref{otherspecies}, we give 
predicted radial and beam-averaged column densities
at $t = 3\times 10^{4}$ yrs
for various species with $N > 10^{13}$ cm$^{-2}$.
The beam-averaged column densities assume a gaussian
beam of full-width at half-max of 15 arsec, though 
the results are insensitive to this assumption.

   \begin{table}
      \caption[]{Predicted Column Densities at $t=3\times 10^{4}$ yrs.}
         \label{otherspecies}
         \begin{tabular}{lrr}
            \hline
            Species & $N_{\mathrm{radial}}(\mathrm{X})$ & 
	    $N_{\mathrm{beam}}(\mathrm{X})$ -- $15 ''$ \\ 
            \hline
     OI & 4(18) & 3(18) \\
     N  & 3(16) & 2(16) \\
     S  & 1(16) & 7(13) \\
     NO & 1(16) & 5(15) \\
     OH & 9(15) & 1(15) \\
     SO & 4(15) & 2(13) \\
     H$_{2}$CS & 2(15) & 1(13) \\
     C$_{3}$H & 2(15) & 2(15) \\
     C$_{4}$H & 1(15) & 1(15) \\
     C$_{3}$H$_{2}$ & 7(14) & 7(14) \\
     CH$_{3}$OCH$_{3}$ & 6(14) & 7(12) \\
     CHOOH & 4(14) & 1(14) \\
     NH$_{2}$ & 3(14) & 2(13) \\
     CH$_{2}$CO & 2(14) & 2(14) \\
     CH$_{3}$ & 2(14) & 6(12) \\
     NH & 1(14) & 3(12) \\
     H$_{3}$O$^{+}$ & 8(13) & 2(13) \\
     CN & 5(13) & 5(13) \\
     OCN & 5(13) & 1(13) \\
     NS & 5(13) & 9(10) \\
     C$_{2}$S & 4(13) & 3(13) \\
     HS$_{2}$ & 4(13) & 1(12) \\
     C$_{6}$H & 4(13) & 4(13) \\
     C$_{3}$H$_{3}$ & 3(13) & 2(13) \\
     CCN & 3(13) & 3(13) \\
     H$_{2}$C$_{3}$ & 3(13) & 2(13) \\
     CH$_{3}$OH$_{2}^{+}$ & 2(13) & 2(11) \\
     HNO & 2(13) & 1(13) \\
     CH$_{3}$CHO & 2(13) & 7(12) \\
            \hline
     \end{tabular}\\
$^{\mathrm{ }}$ { } $a(b)$ means $a \times 10^{b}$\\
$^{\mathrm{ }}$ { } All column densities given in cm$^{-2}$\\
   \end{table}

\subsection{Implications for mantles and mantle destruction}

\begin{figure}
    \resizebox{\hsize}{!}{\includegraphics{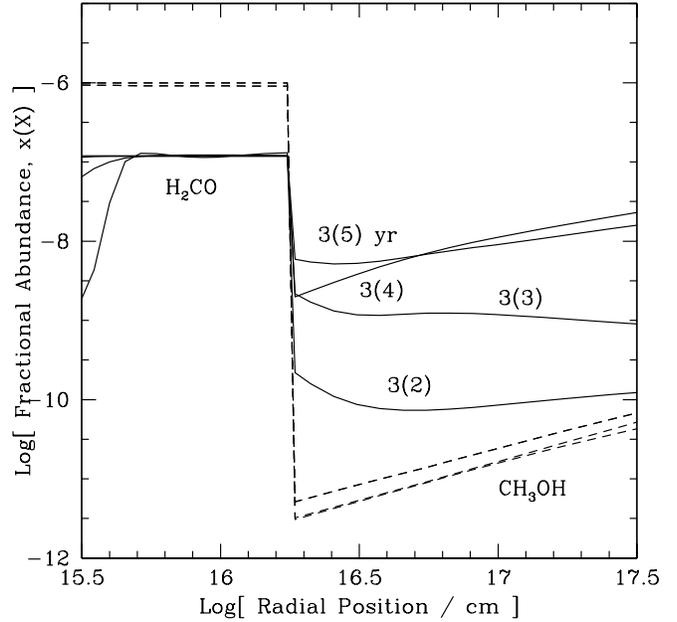}}
      \caption[]{The fractional abundance of H$_{2}$CO and CH$_{3}$OH 
            throughout the envelope for various times.  The
	    curves are labeled by the time in years, where
	    $a(b) = a \times 10^{b}$. 
	    Note that both species are initially depleted from
	    the gas phase for $T \leq 100$ K. 
              }
         \label{fig.xh2coxch3oh.vsr}
   \end{figure}

Finally, although we do not explicitly consider grain-surface
chemistry, it is worthwhile
to discuss the implications our results have on the
grain mantles, and grain-surface chemistry.
Two observed species that may form on grain mantles
are H$_{2}$CO and CH$_{3}$OH.
In Fig. \ref{fig.xh2coxch3oh.vsr} we plot the fractional 
abundance of H$_{2}$CO and CH$_{3}$OH throughout the
envelope at various times.  

%

When low-temperature depletion is assumed, the 
column densities for H$_{2}$CO and CH$_{3}$OH are generally close  
to the observations.  
In particular, while van
der Tak, van Dishoeck \& Caselli (\cite{vvc}) report column densities
of $N($H$_{2}$CO$) = 8 \times 10^{13}$ cm$^{-2}$ and
$N($CH$_{3}$OH$) = 1.2 \times 10^{15}$ cm$^{-2}$ respectively,
the model predicts a CH$_{3}$OH column density 
about 6 times lower, and an H$_{2}$CO column density about
5 times higher.  

On the other hand, detailed radiative transfer modeling 
by van der Tak, van Dishoeck, \& Caselli (\cite{vvc}) 
suggests that the observed lines are consistent with a 
uniform H$_{2}$CO abundance of $4 \times 10^{-9}$,
and a CH$_{3}$OH abundance of $2.6 \times 10^{-9}$ for 
$T \leq 100$ K, and $8 \times 10^{-8}$ for $T \geq 100$ K. 
For comparison the predicted abundances for these species  
are shown in Fig. \ref{fig.xh2coxch3oh.vsr}. 
The H$_{2}$CO abundance in the cool exterior is consistent
with the inferred abundance, while the  
abundance in the warm interior is predicted to be significantly
higher and decreases only slowly with time.  
The CH$_{3}$OH, on the other hand, does not fit the 
inferred abundances very well.  While there does exist a 
``jump'' as suggested by van der Tak, van Dishoeck, \& Caselli
(\cite{vvc}), the abundances predicted by the model are significantly
too low in the exterior and too high in the interior.

%
   \begin{figure}
    \resizebox{\hsize}{!}{\includegraphics{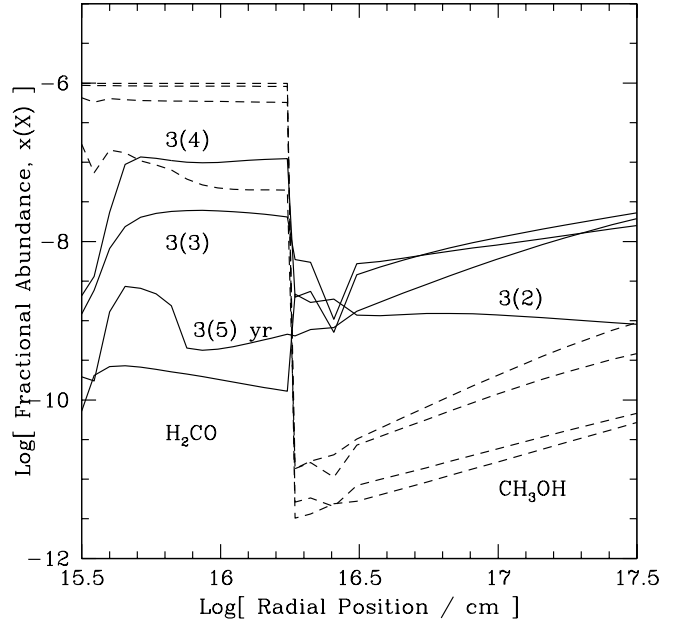}}
      \caption[]{The fractional abundance of H$_{2}$CO and CH$_{3}$OH 
            throughout the envelope for various times, after
	    a hypothetical non H$_{2}$-dissociative heating event.  The
	    curves are labeled by the time in years, where
	    $a(b) = a \times 10^{b}$. 
	    Note that both species are initially depleted from
	    the gas phase for $T \leq 100$ K. 
              }
         \label{fig.xh2coxch3oh.vsr.afterflare}
   \end{figure}

For comparison, we have also run models where the abundances of 
H$_{2}$CO and CH$_{3}$OH are initially undepleted for
$T \leq 100 ~ \mathrm{K}$.
In models where the cold initial abundances of CH$_{3}$OH are set equal
to the hot initial abundances
only CH$_{3}$OH, C$_{3}$H$_{3}$, and CH$_{3}$OCH$_{3}$ show
column density differences of a factor of three or more.
When we adopt this initial abundance, the
column density of CH$_{3}$OH increases to 
$4.4 \times 10^{15}$ cm$^{-2}$, a value approximately 4 times
larger than the observations.
These results suggest that while we can reproduce the
CH$_{3}$OH column density with some accuracy, simple gas-phase
chemistry alone cannot reproduce the apparent details of the
CH$_{3}$OH abundance distribution.  As a result, it appears
that CH$_{3}$OH can be strongly affected by grain surface
chemistry, also in the cooler regions.   

Allowing a cold H$_{2}$CO abundance equal to the warm
abundance changes the predicted column density by only
a factor of two, with only minor differences for all
other species. 
This implies that
only observations of high-enough spatial resolution to 
differentiate between the warm and cold phases, or
use of high-excitation lines will
be able to best determine the nature of H$_{2}$CO formation.

It is also interesting to consider the fact
that submillimeter observations suggest that the 
abundances of warm H$_{2}$CO and CH$_{3}$OH are 
factors of 100 -- 1000 below the solid state abundances.
In our models, we find that the ratio of warm to
cold H$_{2}$CO for a 15 arcsecond beam is
$60 \leq [N($H$_{2}$CO$)]_{T \geq 100}
/[N($H$_{2}$CO$)]_{T \leq 100} \leq 500$ for
$3 \times 10^{4} \leq t(\mathrm{yrs}) \leq 10^{5}$.
On the other hand, for CH$_{3}$OH we find that
$1 \leq [N($CH$_{3}$OH$)]_{T \geq 100}
/[N($CH$_{3}$OH$)]_{T \leq 100} \leq 2$ for
$3 \times 10^{4} \leq t(\mathrm{yrs}) \leq 10^{5}$.
This confirms the previous suggestions that 
that gas-phase chemistry may perhaps
dominate grain-surface chemistry in the production
of H$_{2}$CO, while 
there must be some other (presumeably grain-surface) pathway
to the production of CH$_{3}$OH.

As a final note, we have considered the effect of a
heating event (as proposed in Sect. 4.4) in which H$_{2}$
is not dissociated on the H$_{2}$CO
and CH$_{3}$OH chemistry.  The results are shown in 
Fig. \ref{fig.xh2coxch3oh.vsr.afterflare}.  
As can be seen, the later-time abundances are more consisent
with the results inferred by van der Tak, van Dishoeck, 
\& Caselli (\cite{vvc}).  In particular, the 
CH$_{3}$OH abundance in the interior is in the range of 
$5 \times 10^{-8} \leq x($CH$_{3}$OH$) \leq 6 \times 10^{-7}$,
while in the exterior the abundances can reach as high as 
$0.3 - 1 \times 10^{-9}$.  Likewise, the
H$_{2}$CO abundance for $3 \times 10^{4} \leq t($yrs$) \leq
3 \times 10^{5}$ is in the range 
$10^{-7} \geq x($H$_{2}$CO$) \geq 3 \times 10^{-10}$. While  
not conclusive, this
brackets the inferred H$_{2}$CO abundance of $4 \times 10^{-9}$
nicely. 
If further suggestions of a heating event are found, 
it may be useful to re-visit these data for comparison
with observations as they may provide a gas-phase 
mechanism for the production of H$_{2}$CO and CH$_{3}$OH.

\section{Time constraints}
Based upon the large amount of observational data for
AFGL 2591 (see Table \ref{molecularobs}), and given 
the time-dependent nature of the reaction network, 
one important test of the physical and chemical 
model would be a determination of the chemical evolution
time of the envelope, consistent with all or most
of the observed species.  This has been proposed and
carried out previously (e.g., Stahler \cite{stahler}; 
Millar \cite{millartime1990};
Helmich et al. \cite{helmichetal1994};
Hatchell et al. \cite{hatchelletal1998}) in single-point models
of dense cloud cores, with some success.

%
   \begin{figure}
    \resizebox{\hsize}{!}{\includegraphics{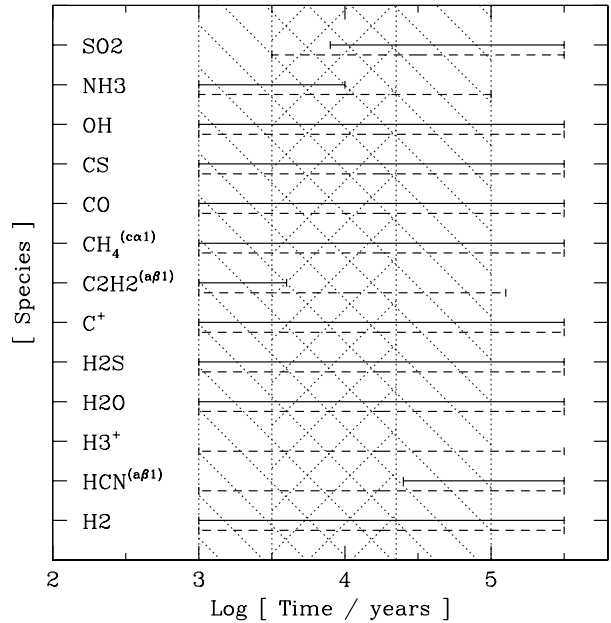}}
      \caption[]{A comparison of the predicted and observed
                 abundances and column densities for the 
                 species listed in Table \ref{molecularobs},
                 and observed in the infrared (see text).
                 The solid lines correspond to agreement between
                 the models and observations within a factor of 3,
                 and the dashed-lines to within a factor of 10.
                 The species are listed, with notes on the
                 observational fits given as parentheses as in
                 Table \ref{molecularobs}.  The two shaded regions
                 denote the regions of potential and preferred fit
                 between the model and the observations(see text).  Notice
                 the agreement with Fig. \ref{fig.time.submm}. 
              }
         \label{fig.time.ir}
   \end{figure}
%

%
   \begin{figure}
    \resizebox{\hsize}{!}{\includegraphics{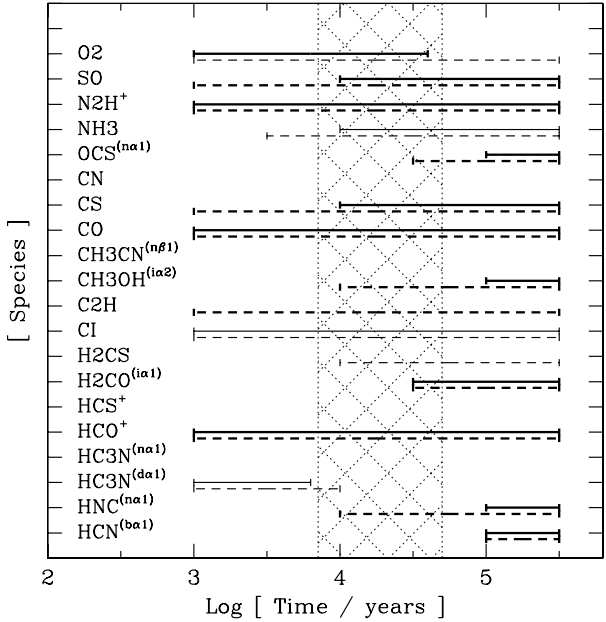}}
      \caption[]{A comparison of the predicted and observed
                 abundances and column densities for the 
                 species listed in Table \ref{molecularobs},
                 and observed in the submillimeter (see text).
                 The solid lines correspond to agreement between
                 the models and observations within a factor of 3,
                 and the dashed-lines to within a factor of 10
		 (see text for details).
                 The species are listed, with notes on the
                 observational fits given as parentheses as in
                 Table \ref{molecularobs}.  The shaded region
                 denote the regions of preferred fit
                 between the model and the observations.  Notice
                 the agreement with Fig. \ref{fig.time.ir}.
              }
         \label{fig.time.submm}
   \end{figure}
%

In our case, we determine the time-dependent 
fractional abundances and column densities for each of 
the species observed in Table \ref{molecularobs}.  
As discussed in Sect. 2.3, we divide the data into two sets:
those for which infrared absorption measurements have been made, and
those for which submillimeter emission measurements
have been made. 

In Figs. \ref{fig.time.ir} and \ref{fig.time.submm} we plot
the approximate time ranges over which the listed species match the
observed data.  In both figures, the solid lines represent
agreement to within a factor of 3, while the dashed lines
represent agreement to within a factor of 10.
These limits can be considered good and acceptable levels
of agreement respectively
(see e.g., Millar \& Freeman \cite{millarfreeman}, and
Brown \& Charnley \cite{browncharnley}
In 
both figures, the lower limits to the chemical
evolution time are capped at $10^{3}$ years.
As in Table \ref{molecularobs}, 
observational data are appended to each species name.  
We include all species from Table \ref{molecularobs}, except
for CO$_{2}$.  As discussed in Sect. 4.4, there are significant
discrepancies and questions about the gas-phase chemistry
of CO$_{2}$, and as such we have treated it separately
in that section.
For species other than CO$_{2}$, we include
the data that is relevant to comparison with our
model 
(i.e., the most reliable components).  

In Fig. \ref{fig.time.ir}, we can see a wide variation of
possible times when constraining the models by the infrared data.
We place two limits on the evolution:
a wide range of $10^{3} \leq t(\mathrm{yrs}) \leq 1 \times 10^{5}$, 
and a preferred limit of 
$3 \times 10^{3} \leq t(\mathrm{yrs}) \leq 2.5 \times 10^{4}$.
These regions are identified by the shading in Fig. \ref{fig.time.ir}.

For comparison, in Fig. \ref{fig.time.submm} we plot the 
constraints on the time for the submillimeter data.  
As discussed in Sect. 2.3, results based upon 
sophisticated, self-consistent, radiative transfer modeling
are given slightly more weight, and identified by
the bold lines in Fig. \ref{fig.time.submm}.  Also, the
O$_{2}$ upper limit by SWAS is given more weight.  This is done
because in the absence of upper limits for the O$_{2}$ abundance
toward AFGL 2591 in particular, we have used the largest quoted upper limit of
 $N(\mathrm{O}_{2})/N(\mathrm{H}_{2}) \leq 9 \times 10^{-7}$ 
(Goldsmith et al. \cite{swaso2}).
Where radiative-transfer modeling derived abundances are 
not available, we have calculated the appropriate beam-averaged
column densities for comparison with the observations.

Most species in Fig. \ref{fig.time.submm} fit the models
to within an order of magnitude of the observational data.
Those that do not fit are not significant defects, for
a number of reasons.
First, not all data are in disagreement with the models -- 
both OCS and HC$_{3}$N have other observations / reductions 
which do agree with the models.
Second, the discrepancies can be understood on a case-by-case
basis.  For instance, the chemistry and reaction rates of OCS
are only poorly understood at best (Millar, private communication).  
It is interesting to note, however, that the
radial OCS column density matches the observed column density.
Also, we expect the potential difficulties with species
related to HCN (such as HC$_{3}$N and CH$_{3}$CN) 
as our model does not probe
the complete region over which significant HCN production
may be important (see Sect. 4.2 above).
In the case of HCS$^{+}$, the abundance is strongly affected
by enhancements of CS abundance at temperatures of $10-20$ K
(Helmich \cite{helmichthesis}),
lower than all but our outermost temperature, signifying that the 
AFGL 2591 envelope may be more extended than we have assumed.
In a similar fashion, CN is strongly influenced by 
UV radiation from a PDR (Helmich \cite{helmichthesis}),
a radiation source not considered in our model.  

The age-constraints implied by the results in Fig. \ref{fig.time.submm}
suggest chemical evolution times in the range
$7 \times 10^{3} \leq t(\mathrm{yrs}) \leq 5 \times 10^{4}$, with
a strong preference for $t \sim 3 \times 10^{4}$ years.  
These constraints are shown by the shaded region in the figure.

\begin{figure}
    \resizebox{\hsize}{!}{\includegraphics{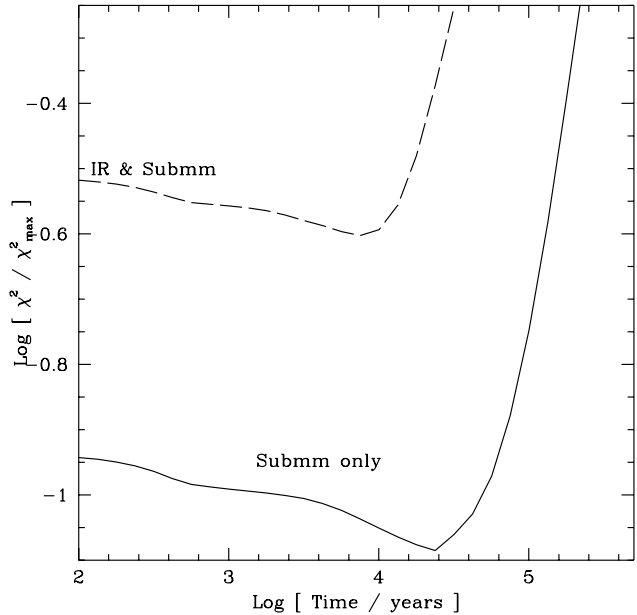}}
      \caption[]{The quality of the model fit to the observations
                 as a function of time.  Here we plot the
                 $\chi^2$ as defined in the text, normalized to the
                 maximum value for the times considered.  The solid
                 line shows the chi-squared for the submillimeter
                 data only, while the dashed-line shows the results
                 when both the submillimeter and infrared data 
                 are included.
              }
         \label{fig.chisq.vst}
   \end{figure}

In order to attempt to quantify this result, in 
Fig. \ref{fig.chisq.vst} we plot the  
chi-squared value between the models and the observations,
normalized to the maximum chi-squared in the entire time evolution.
The $\chi^2$ is defined by 
$\chi^2 \equiv 
\Sigma_{i} w_{i}^2 \times [y_{\mathrm{model},i}-y_{\mathrm{obs},i}]^2/
\sigma_{i}^2$.  Here we include the weight from Table
\ref{molecularobs} as $w$, and assume uncertainties of a factor
of 5 in the observations for all species except O$_{2}$, for
which we assume a factor of 2 uncertainty as we already 
have adopted the highest observed upper limit from 
Goldsmith et al. (\cite{swaso2}). 
These results are generally consistent with those of 
Figs. \ref{fig.time.ir} and \ref{fig.time.submm}, both in
terms of the preferred times as well as the relatively
lower level of constraint provided by the IR data.  In particular,
while times of up to $\sim 10^5$ yrs may be possible, there
appears to be a preference for somewhat lower times near
$\sim 3 \times 10^4$ yrs.  

It is interesting and reassuring that the chemical evolution
times from both the infrared and submillimeter data 
provide similar results.  This is especially true as 
they probe such different regions of the envelope.
Also, the nearly simultaneous agreement in time between such
different species, with transitions arising throughout the
envelope, and observed with a range of ground- and space-based
instruments provides significant support to the proposed
chemical, physical, and thermal structure of the envelope.

%
%

\section{Conclusions}
We have constructed detailed thermal and gas-phase
chemical models for AFGL 2591 based upon the physical model
of van der Tak et al. (\cite{vdt1999}; \cite{vdt2000}). 
These models were used to probe the validity of the proposed
physical structure, as well as study the chemical evolution of the source.  

In particular, we find that:
\begin{enumerate}

\item Pseudo-time dependent modeling of the chemistry
as a function of depth is a good probe of the physical and
thermal structure of the source,  due to the density
and temperature dependence of the chemistry.  

\item Care must be exercised when drawing conclusions from
observations, as infrared absorption measurements usually better probe
the warm interior of the envelope while submillimeter emission data
often better probe the cool exterior.  While ``average abundances''
[$x_{\mathrm{avg}} \equiv N(\mathrm{X})/ N(\mathrm{H}_{2})$] can
suggest conditions different from those actually in existence,
detailed radiative transfer modeling are essential to give more
reliable results (Sect. 2 \& 4.1).  This also underscores the
importance of modeling the complete physical, thermal and chemical
structure of the envelope when comparing to observations.

\item Water and CO are stable, even at high temperatures,
except for destruction by cosmic-ray-driven ion-molecule
reactions for $t > 10^{5}$ yrs.  The ice and gas abundances
toward other sources suggest that there may exist a
water destruction mechanism at high temperatures
($> 500$ K).  Such 
a mechanism does not exist in our chemical model, but is
worth further exploration
(Sect. 4.1 \& 4.2).

\item The hydrocarbon and nitrogen chemistry is strongly 
influenced by temperature, with significant production possible
at high ($T \geq 400$ K) temperatures.  Still it appears that
there may exist an as-yet unexplored path to HCN for $T \geq 600$ K
(Sect. 4.2).

\item The sulphur chemistry displays three significant regimes.
In the cool exterior,  
over 50\% of the sulphur exists in CS.  
For $100 \leq T(\mathrm{K}) < 300$, 
SO$_{2}$ contains a majority of the sulphur. At higher temperatures,
atomic sulphur dominates.  While consistent with observations,
these results employ a low gas-phase sulphur abundance.  We cannot
identify the principle sulphur reservoir in the cold gas
(Sect. 4.3).

\item Carbon dioxide may possibly be destroyed by reactions with
molecular hydrogen in impulsive heating events -- 
a reaction which requires further study.
The suggestion that CO$_{2}$ may be destroyed by H has 
problems in that it may destroy too much CH$_{4}$ in the interior
and produce too much atomic and molecular oxygen
(Sect. 4.4).

\item The cosmic-ray ionization rate in AFGL 2591 is important.  It
affects the ion abundances as well as the formation of HCN.  These
processes fix the cosmic ray ionization rate to better than a factor
of three.  Ions produced by cosmic rays can also effectively destroy
H$_{2}$O on timescales of $> 10^{5}$ years (Sect. 4.5).

\item We produce simulated column densities
of predicted abundant species for comparison with
future observations (Table \ref{otherspecies}).
We predict, among other things, abundant atomic oxygen and nitrogen --
even though oxygen can be shuttled to water and
O$_{2}$ while nitrogen is at most 1 \% atomic with the 
majority in N$_{2}$.  The temperature variation within the source can have
a large affect on oxygen, and hence on the rest
of the chemistry.  
(Sect. 4.6).  

\item While it is possible for the observed H$_{2}$CO
abundance to be reproduced by gas-phase chemistry, 
the same is not true for CH$_{3}$OH.
This suggests that some other processes are important for CH$_{3}$OH, 
including grain-surface chemistry and/or the
potential of a heating event within the envelope
(Sect. 4.7).

\item It is possible to use detailed chemical modeling to constrain 
the chemical age of AFGL 2591.  
We find $7 \times 10^{3} \leq t(\mathrm{yrs}) \leq 5 \times 10^{4}$,
with a strong preference for $t \sim 3 \times 10^{4}$ years
(Sect. 5).

\item The agreement of our results with the line data lend
significant further confirmation to the physical model
proposed by van der Tak et al. (\cite{vdt2000}), and the
chemical structure and evolution proposed here. 

\end{enumerate}

\begin{acknowledgements}
      We are grateful to the anonymous referee for comments
      and discussions that helped improve the presentation.
      This work was partially supported under
      grants from The Research Corporation (SDD), 
      and the Netherlands Organisation for Scientific Research
     (NWO) through grant 614-41-003 (AB, FvdT),
      and a NWO bezoekersbeurs. Astrochemistry at Leiden is
     supported through an NWO Spinoza award (EvD).
\end{acknowledgements}

\end{document}